\documentclass{article}

\usepackage[utf8]{inputenc}
\usepackage{amsmath}
\usepackage{amssymb}
\usepackage{graphicx}
\usepackage[a4paper, left=27.0mm, right=27.0mm, top=20.3mm, bottom=30.5mm]{geometry}
\usepackage{float}
\usepackage[colorlinks]{hyperref}
\usepackage{caption}
\usepackage{empheq}

\usepackage[sorting=none, backend=bibtex, style=numeric, maxbibnames=20]{biblatex}
\title{Li\'enard-Wiechert Potentials due to a `classically' spinning point-charge in Non-Relativistic arbitrary motion}
\author{Nikhil D. Hadap\\Bhabha Atomic Research Centre\\Trombay, Mumbai- 400085, India}
\date{\today}

\begin{document}

\maketitle

\begin{abstract}
Li\'enard-Wiechert potentials have been derived for a moving and `classically’ spinning point-charge; assuming it to be a small rigid charged-sphere in combined non-relativistic translational and rotational motion, and subsequently reducing its dimensions to `point-particle’ limit. The paper demonstrates that when the effect of rotation were taken into account, together with causality, expressions for the LW potentials accompany additional correction terms that contain spin-angular-momentum (or simply the `classical-spin') of the point-charge.
\end{abstract}

\section{Introduction} \label{intro}

The principle of `causality' states that no effect can occur outside the future light cone of its `cause'. \cite{rohr} \bigskip

However, because ``classical physics'' lacks a ``arrow of time'' (except ``law of entropy'' for macroscopic systems), 'retardation' in classical field theories is introduced by hand to obtain causal solutions. \cite{rohr} ``Retarded Potentials'' are such a set of electromagnetic potentials that are used to obtain causal solutions in the field theory of classical electrodynamics. \bigskip

In classical electrodynamics, retarded potentials appear because ``electromagnetic news travels with speed of light'', \cite{grif} which causes the information from different parts of a dynamic charge-configuration to reach the observer at different times; and forces the volume integration of charge and current densities to be evaluated at different times. \cite{grif} \bigskip

Such a set of `causal' potentials are known as the ``Retarded Potentials'', given by\cite{grif} \cite{jak}:

\begin{equation} \label{1}
\left.
\begin{matrix}
\Phi ( \textbf{R}, t ) & \ = \ &
\dfrac{ 1 }{ 4 \pi \epsilon_0 } \iiint_V { \dfrac{ \rho ( \textbf{r'}, t_{ret} ) }{ R } dV' } \\
{} & {} & {} \\
\textbf{A} ( \textbf{R}, t ) & \ = \ &
\dfrac{ \mu_0 }{ 4 \pi } \iiint_V { \dfrac{ \textbf{ j } ( \textbf{r'}, t_{ret} ) }{ R } dV' }
\end{matrix}
\right\}
\end{equation} \medskip

Where, $ \rho ( \textbf{r'}, t_{ret} )$ and $ \textbf{j} ( \textbf{r'}, t_{ret} ) $ being the charge and current densities that prevailed at point $ \textbf{r'} $ and retarded time $ t_{ret} $, respectively.  \bigskip

For a point charge, the charge and current density functions $ \rho ( \textbf{r'}, t_{ret} ) $ and $ \textbf{j} ( \textbf{r'}, t_{ret} ) $ are either replaced by direc-delta functions \cite{jak}, or they are evaluated for an extended charge distribution in the limit of its size approaching to zero.\cite{grif} \bigskip

The potentials thus obtained for the moving point charge are given by\cite{grif} \cite{jak} :

\begin{equation} \label{2}
\left.
\begin{matrix}
\Phi ( \textbf{R}, t ) & \ = \ &
\dfrac{ q }{ 4 \pi \epsilon_0 } \dfrac{ c }{( R c \ - \textbf{R} \cdot \textbf{v} )} &
\ = \ \dfrac{ q }{ 4 \pi \epsilon_0 R } \dfrac{ 1 }{( 1 \ - \boldsymbol{\hat n} \cdot \boldsymbol{ \beta } )} \\
{}& {}& {}& {}\\
\textbf{A}( \textbf{R}, t ) & \ = \ &
\dfrac{ q }{ 4 \pi \epsilon_0 c } \dfrac{ \textbf{v} }{( R c \ - \textbf{R} \cdot \textbf{v} )} &
\ = \ \dfrac{ q }{ 4 \pi \epsilon_0 R c } \dfrac{ \boldsymbol{ \beta } }{( 1 \ - \boldsymbol{\hat n} \cdot \boldsymbol{ \beta } )} \\
{}& {}& {}& {}\\
{}& \ = \ & \dfrac{ \textbf{v} }{ c^2 } \Phi ( \textbf{R}, t) & {}
\end{matrix}
\\ \right\}
\end{equation} \medskip

These are famously known as Li\'enard-Wiechert (LW) potentials, developed in part by Alfred-Marie Liénard in 1898 and independently by Emil Wiechert in 1900.  \bigskip

In 4-vector form, with flat space-time (Minkowski) metric being $ \eta_{ \alpha \beta } \ = \ diag \left[ -1; 1; 1; 1 \right] $, LW 4-potentials may be written as:

\begin{equation} \label{3}
{ A^\mu }( R^\mu) \ = \ \dfrac{ q }{ 4 \pi \epsilon_0 c^2 } \ \dfrac{ u^\mu }{ \left( - R_\mu u^\mu \right) }
\end{equation} \medskip

Where: $ R^{\mu} \ = \ \left[ c t_{ret} ; \ \textbf{R} \right] $ is the instantaneous 4-distance connecting the `retarded' position of the point-charge, at time $ t \ - t_{ret} $ ), to the distant observer at present time $ t $. And $ u^{\mu} \ = \left[ c ; \ \textbf{v} \right] $ being the instantaneous 4-velocity of the point-charge at the `retarded' position. \bigskip

Since equations (\ref{2})  and  (\ref{3}) make no reference to the dimensions of the charge, they do hold equally for extended charges as well, provided their dimensions are  negligible in comparison to the distances at which the fields are being evaluated. \cite{pan} \cite{van} \bigskip

In classical electrodynamics the only thing known about a point charge (e.g. electron) is that it has a specific `total' charge, and any calculations of its electromagnetic fields do not take into account how this charge is distributed geometrically within it. \cite{pan} \bigskip

Further, it is impossible to assume that the charge could have `zero' physical extent without introducing various mathematical divergences. However, most field characteristics are independent of the charge's physical geometry and dimensions, as long as they are tiny in comparison to the distances at which the fields are being observed. \cite{pan} \bigskip

In this paper, we assume that our point-charge has a finite size; though, negligible compared to the distance from which its fields are being observed. In other words, it appears like a `point' to the distant observer. Thus, in this paper mathematically we mean the ``point-particle limit'' as: ``dimensions are sufficiently close to zero'', rather than ``dimensions are zero''. \bigskip

The purpose of this paper is to extend equations (\ref{2}) for a moving point-charge, that is also rotating about its center-line; or in other words ``spinning classically''. \bigskip

Special relativity further limits the theoretical size of a `spinning' `point-particle' to constraint the tangential velocities of its surface below the speed of light, though the same is not in the scope of this paper, which is limited to only “non-relativistic” motion, including rotation. \bigskip

In this work `classical-spin' (``Spin Angular-Momentum'' of a point-charge) has been considered a `bulk' `invariant' observable property of the point-charge independent of whatever geometry and velocities its different parts might have inside; much like the `total-charge' which is independent of whatever geometry and charge-densities its parts might have inside. \bigskip

Arakelyan et al. \cite{ark}, evaluated LW potentials for ``pseudo-classical spinning point-particle'' in 2000. However, in their work spin terms appear alongside $ 1/ { R ^2 } $; which is essentially due to the fact that in their work ( and their earlier cited works \cite{ark}) `spin' was described in terms of ``magnetic dipole-moment'' that appeared from ``Multi-pole expansion'' of potentials. \bigskip

In contrast, our work shows `spin' as the fundamental quantity: ``spin angular-momentum''. Also, In our work, spin terms appear alongside $ 1/ R $ as well. \bigskip

To further comprehend, we assume that our `spinning' point-charge is spherically symmetrical. Though, several other configurations may be feasible, we believe that as long as the charge's physical extent is negligibly small, the outcome will be the same (or of the same order of magnitude). \bigskip

Griffiths \cite{grif} and Lorentz \cite{lor} performed such calculations to determine the `Self-Force' for a point charge. While Lorentz evaluated the self-force for a spherical charge, Griffiths demonstrated it for `dumbbell' configuration in which the total charge was divided into two lobes separated by a short distance. However, in `point-particle' limit both the cases produced the same result. \bigskip

Lastly, we approximate our charge to be `rigid'; so that all parts of the charge would be co-moving with the center and accelerate with the same amount with time, in the observer's rest-frame. This assumption greatly simplifies our mathematics, because in the observer's rest-frame the charge would retains its spherical shape (though, apparently it would looks deformed due to retardation effects.) \bigskip

Also, as our charged sphere has negligible physical extent, it would perceive the force-field, through which it is moving, as nearly `uniform'; thus, causing the same acceleration for all the points within the sphere. \bigskip

In short, our work seeks to demonstrate `retardation' (not `relativistic') effects measured by a distant observer, due to a non-relativistically moving and rotating (classically spinning), rigid and negligibly small charged sphere. \bigskip

The organization of this paper is as follows: \bigskip

Section- \ref{geo_rod} demonstrates how a body's acceleration also significantly distorts the apparent geometry that a distant observer perceives. \bigskip

Section- \ref{rigid} analyses our ``rigid body assumption'' and fixes further constraints on non-relativistic motion. \bigskip

Section- \ref{geo_spher} sums up the postulate for a spinning, rigid, spherical volume charge distribution in arbitrary motion. \bigskip

Section- \ref{derive} continues with the derivation of LW 4-potentials for a non-relativistically pinning and moving point-charge; and demonstrates the presence of extra  terms that contain `classical-spin' of the point-particle. \bigskip

Section- \ref{interp} attempts to analyze significance of the obtained extra terms. \bigskip

Section- \ref{valid} confirms the validity of obtained expression. \bigskip

Section- \ref{conclu} is the conclusion, followed by references. \medskip

\section{Electromagnetic news received from an Accelerated Object:} \label{geo_rod}

``Electromagnetic news travels with speed of light $ c $''. Thus, in non-static case, it is not the status of the source ``right now'' that matters, but rather its condition at some earlier time $ t_{ret} $ when the `message' departed.  Since, this message must travel a distance $ R $ to the observer, it gets delayed by $ R/ c $.\cite{grif} This `earlier' time $ t_{ret} $ is called ``Retarded Time''. \bigskip

When a non-static charge-configuration is viewed by a distant observer, it rather presents a distorted picture of total charge and current. This is because the retardation $ t_{ret} = t - R/ c $ obliges evaluation of charge and current densities $ \rho $ and $ \textbf{j} $, respectively, for different parts of the configuration at different times. \cite{grif} \bigskip

This is a purely geometrical effect.\cite{grif} \bigskip

\begin{figure}[H]
\centering
\includegraphics[width=6cm]{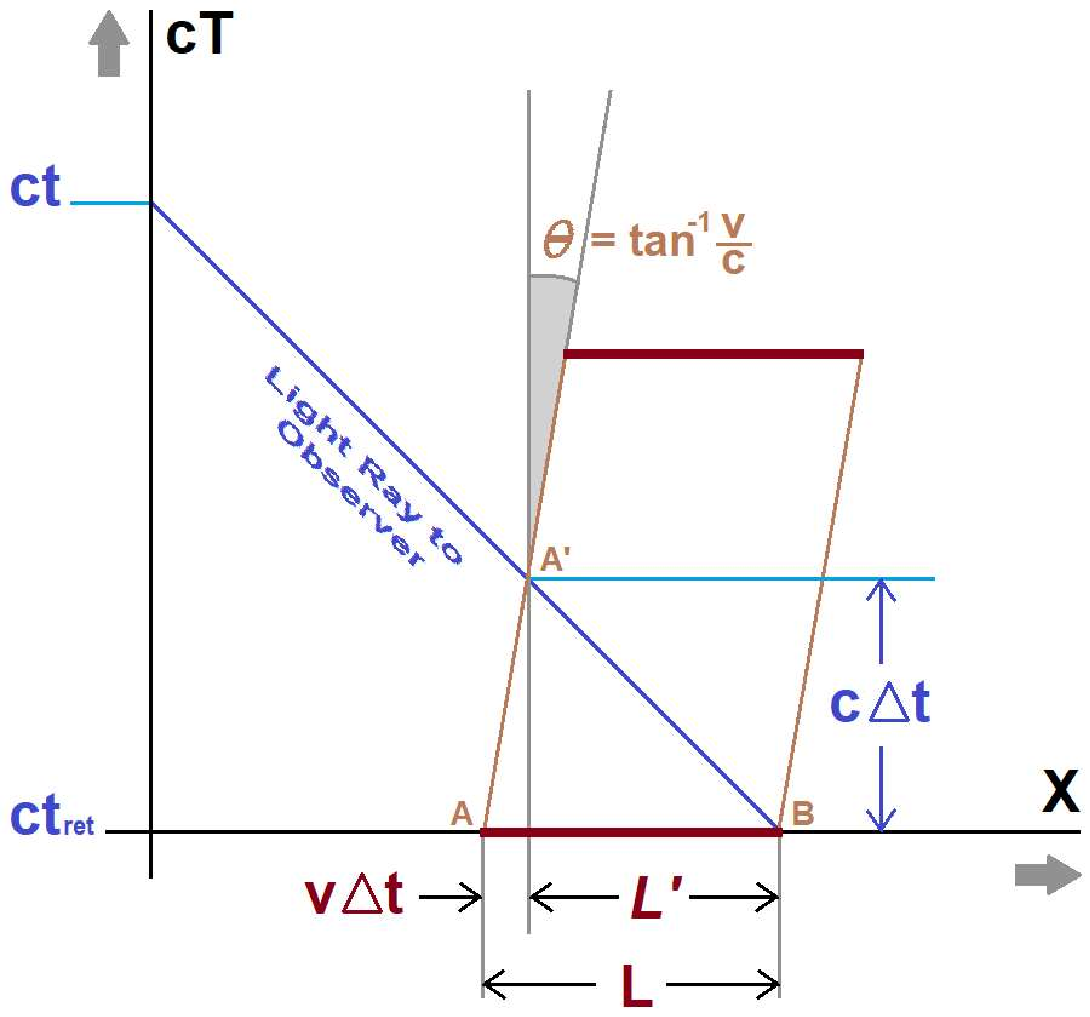}
\caption{(Color Online) A rigid rod moving in 1-dimension with constant velocity.}
\label{fig:1}
\end{figure}

Considering a rigid rod $ A B $ of length $ L $ (measured in observer's rest frame) lying along the X-axis and moving in the direction of positive X-axis with constant velocity $ v $, as in spacetime diagram of figure-\ref{fig:1}.

The `knowledge' about the rod that the distant observer at $ x = 0 $ perceives, is due to `information' that he receives from the rod at ``one instant of time''. However, because of retardation, by the time the light-ray from far end $ B $ meets the opposite end $ A $ on its journey to the distant observer, it will have moved to $ A' $, by amount $ v \Delta t $. While $ \Delta t $ being the time it takes the light-ray from one end of the moving rod to other, ``as viewed by the observer''. \bigskip

Since, the velocity is constant, locus of any point on the rod, connecting its retarded-position at $ T = t_{ret} $ to its present position $ T = t_{ret} + \Delta t $ (e.g. $ A A' $ ), will be a straight line on the spacetime diagram. The slope of this line is determined by the magnitude of the velocity as $ \tan \theta = v / c = /beta $. \cite{mer}. \bigskip

Therefore, the apparent length perceived by the distant observer comes out to be $  L' = L - v \Delta t $. \bigskip

Solving for $ L' $ and $ \Delta t $, with $ L' = c \Delta t $, we get:

\begin{equation} \label{4}
\dfrac{ L'}{ L } \ = \ \dfrac{ c }{( c \ + v )} \ = \ \dfrac{ 1 }{( 1 \ + \beta )}
\end{equation} \medskip

Which may be generalized for motion in arbitrary direction as:

\begin{equation} \label{5}
L' \ = \ \dfrac{ L }{( 1 \ - \boldsymbol{ \hat n } \cdot \boldsymbol{ \beta } )}
\ = \ \left( \dfrac{ c }{ c \ - \boldsymbol{ \hat n } \cdot \textbf{ v } } \right) L
\ = \ \left( \dfrac{ R c }{ R c \ - \textbf{ R } \cdot \textbf{ v } } \right) L
\end{equation} \medskip

While, $ \textbf{R} $ is the vector distance pointing from the particle's `retarded' position to the observer's `current' position; and $ \boldsymbol{ \hat n } $ is the unit vector in the direction of $ \textbf{R} $. \bigskip

The quantity $ ( 1 \ - \boldsymbol{ \hat n } \cdot \boldsymbol{ \beta }  ) $ or $ ( R c - \textbf{R} \cdot \textbf{v} ) $ or $ - R_{\mu} u^{\mu} $ in denominator appears purely due to the apparent geometry as perceived by the distant observer. \bigskip

Thus, depending on the particle's velocity vector, the observer perceives the retarded charge and current densities that are either in-access or less than the actual. \cite{pan} \bigskip

Nevertheless, this geometrical effect has nothing to do with ``Lorentz Contraction'', as $ L $ is length of the `moving' rod, and its `rest' length is not an issue. \cite{grif} \bigskip

When the same logic is extended to the accelerating rod, the picture gets distorted further. \bigskip

With the assumption that the rod is `rigid', and is acted upon by a uniform force field, the entire rod (i.e. both the ends as well as all the parts of the rod) would accelerate by same amount in a given ``coordinate time difference''. \bigskip

However, the distant observer would see different parts of the rod moving with different velocities. As a result, to the distant observer,  the rod would not appear rigid; but would appear to expand or shrink over time (as long as the acceleration persists), depending on the direction of motion and strength of acceleration. This may be visualized from figure- \ref{fig:2} below:

\begin{figure}[H]
\centering
\includegraphics[width=6cm]{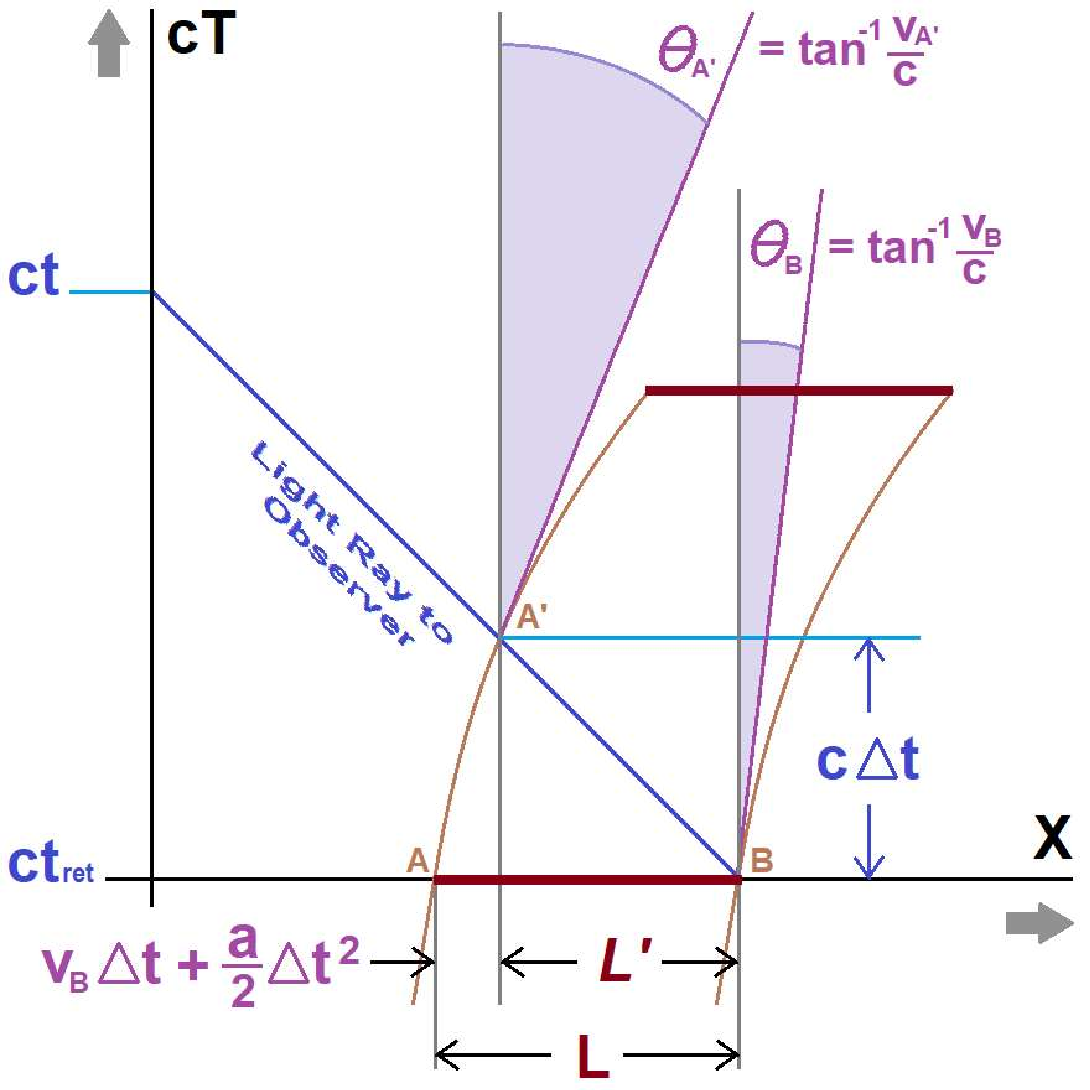}
\caption{(Color Online) Accelerating rigid rod in 1-dimension.}
\label{fig:2}
\end{figure}

From the light-cone carrying the combined information form both the ends, the distant observer would infer that, the far end $ B $ and the near end $ A $ were moving with instantaneous velocities $ v_B $ and $ v_{A'} $ respectively, at the same instance; while $ v_{A'} \neq v_B $. \bigskip

Assuming the velocities and acceleration being small ( $ v_{A'}, v_B << c $; $ a \Delta t << c $ ), non-relativistic (Newtonian) equations of motion may be used. Thus, the distorted length as perceived by the observer now comes:

\begin{equation} \label{6}
L' \ = L \ -  v_B \Delta t \ - \dfrac{ a }{ 2 } \Delta t ^ 2 \ = L \ - \dfrac{( v_B \ + v_{A'} )}{ 2 } \Delta t
\end{equation} \medskip

Though, it seems that for a point-particle with $ L \ \approx 0 $, $ v_{A'} \rightarrow v_B $; and the expressions for LW potentials (\ref{2}) would remain unchanged; however, things are not this simple if the particle is rotating or `spinning'. \bigskip

Jerrold Franklin \cite{jerr} in 2010 shown that, in context of special relativity, for a rigid body to remain 'rigid' under accelerated motion, acceleration in its own frame must vary throughout it's volume to preserve its rest frame dimensions. \bigskip

However, `retardation’ is a phenomenon that is applicable purely in observer’s perspective. ``Retarded-time'' refers to a `past’ coordinate-time that lies in the observer’s rest-frame Therefore, the accelerations experienced by the moving rod within its own frame are irrelevant. \bigskip

In the upcoming section, we will further analyze the validity of “rigid body” assumption.  \medskip

\section{Rigid Body Approximation:} \label{rigid}

`Rigid-body' assumption is not a covariant concept as per relativity; or this would allow forces traveling within the rigid body at infinite speed. \cite{anik} \bigskip

When acceleration or rotation is applied to a body's one part, the information does not reach its other parts instantaneously. Rather, an elastic wave (of speed less than the speed of light) spreads within the body carrying the information to its other parts; causing phase lag among motion of the particles. \cite{anik} \bigskip

Thus, according to relativity, all matter is non-rigid. \bigskip

Considering a non-rigid rod $ A B $ of length $ L_{no-rgd} $, measured in observer's rest frame; lying along the X-axis and moving in the direction of positive X-axis with velocity $ V_1 $, as shown in the space-time diagram of figure- \ref{fig:3} below:

\begin{figure}[H]
\centering
\includegraphics[width=6cm]{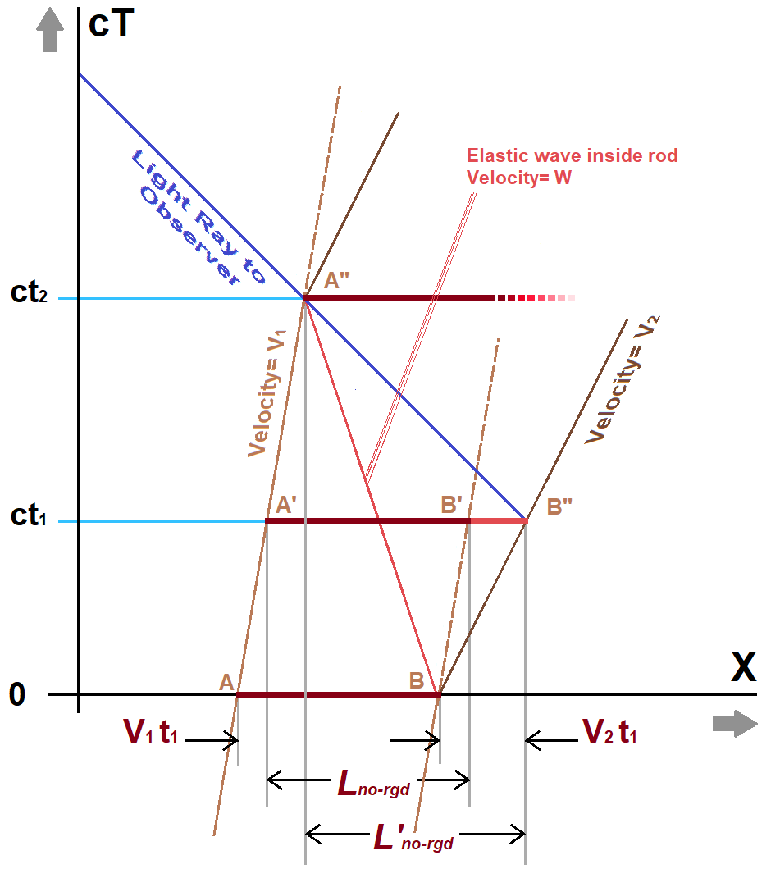}
\caption{(Color Online) Accelerated non-rigid rod.}
\label{fig:3}
\end{figure}

At time $ T = 0 $ the end $ B $ of the rod is dragged in the direction of motion, causing its velocity change to $ V_2 $. The information regarding the change in motion does not reach the other end instantaneously; rather, it is carried by an elastic wave of speed $ W $ that would propagate towards the end $ A $. \bigskip

The end $ A $ continues to move with the earlier velocity $ V_1 $ until the elastic wave encounters it at $ A'' $, at time $ t_2 $. Thus, the rod gets elongated in observer's rest frame. \bigskip

However, as $ W < c $, when the light ray from $ A'' $ reaches the distant observer, it also carries the information from the opposite end when that was at $ B'' $, at time $ t_1 $. Thus, the distant observer perceives the rod as elongated with length $ {L'}_{no-rgd} \ = L_{no-rgd} \ + ( V_2 – V_1 ) \ t_1 $. \bigskip

Also, as $ {L'}_{no-rgd} \ = c \ ( t_2 – t_1 ) $ and $ L_{no-rgd} \ = ( W + V_1 ) \ t_2 $, the apparent elongated length as perceived by the observer can be evaluated as:

\begin{equation} \label{7}
\dfrac{ {L'}_{no-rgd} }{ L_{no-rgd} } \ = \dfrac{ c }{ ( C \ + V_2 ) } \ \left( \dfrac{ W \ + V_2 }{ W \ + V_1 } \right)
\end{equation} \medskip

The change in velocity $ ( V_2 – V_1 ) $ is not straightaway proportional to the applied force; but  is also affected by the elastic resistance offered by the rod's material. Further, in the end, the rod may either retain the elongation, or revert to its original length following several to-and-fro elastic wave oscillations between its ends; depending upon the elastic properties of the rod's material. \bigskip

Though, we don't know how fast elastic waves move within the matter of point-charges, e.g. electron; data on “speed of sound” in elements \cite{wiki} suggests that it could be several thousand times the “speed of sound” in air. \bigskip

A comparison of equation (\ref{7}) and (\ref{4}) shows that the non-rigid condition introduces an elastic factor $ ( W \ + V_2 ) / ( W \ + V_1 ) $ to the apparent retarded length. \bigskip

For $ V_1, V_2 << W $, the elastic factor approaches unity and the retarded length of non-rigid rod approaches to that of a rigid-rod. Thus, for very slow motion, elastic deformation would be negligible; and the body may be considered as ``near rigid''; or may be `approximated' as `rigid'. \bigskip

The next section demonstrates the effect of acceleration on the geometry of a spinning rigid sphere. \medskip

\section{Effect of Acceleration and Spin on apparent geometry of a point-sphere:} \label{geo_spher}

Considering a spherical volume charge distribution of total charge $ q $, radius $ r_q $ and spinning with angular-velocity $ \boldsymbol{ \omega_s } $. The center of the charged-sphere is moving in positive Z-direction, with instantaneous velocity $ \textbf{v} $ and acceleration $ \textbf{a} $. \bigskip

For the sake of simplicity, we assume that the charge distribution's spherical shape is in equilibrium with the rotation; that is, it has a spherical shape while rotating with angular velocity $ \boldsymbol{ \omega_s } $ rather than at `rest'. \bigskip

We also assume that the sphere is `near-rigid', which means that all transnational and rotational velocities are significantly lower than the speed of an elastic wave sustained by the sphere's matter. Thus, any elastic deformations are negligibly smaller than the various transnational and angular displacements. \bigskip

Finally, we assume our sphere is small enough, so that the ``force-causing field'' in which it is moving appears `uniform' to it; as a result, every point within it feels the same acceleration in a given ``coordinate time difference''. \bigskip

Considering an infinitesimal volume-element $ r^2 d \Omega dr $ at position $ \textbf{r} $ inside the sphere, with reference to its center, ( $ dr $ is the thickness, and $ d \Omega $ is the solid-angle that the volume-element subtends at the center of the sphere) as shown in figure- \ref{fig:4}:

\begin{figure}[H]
\centering
\includegraphics[width=10cm]{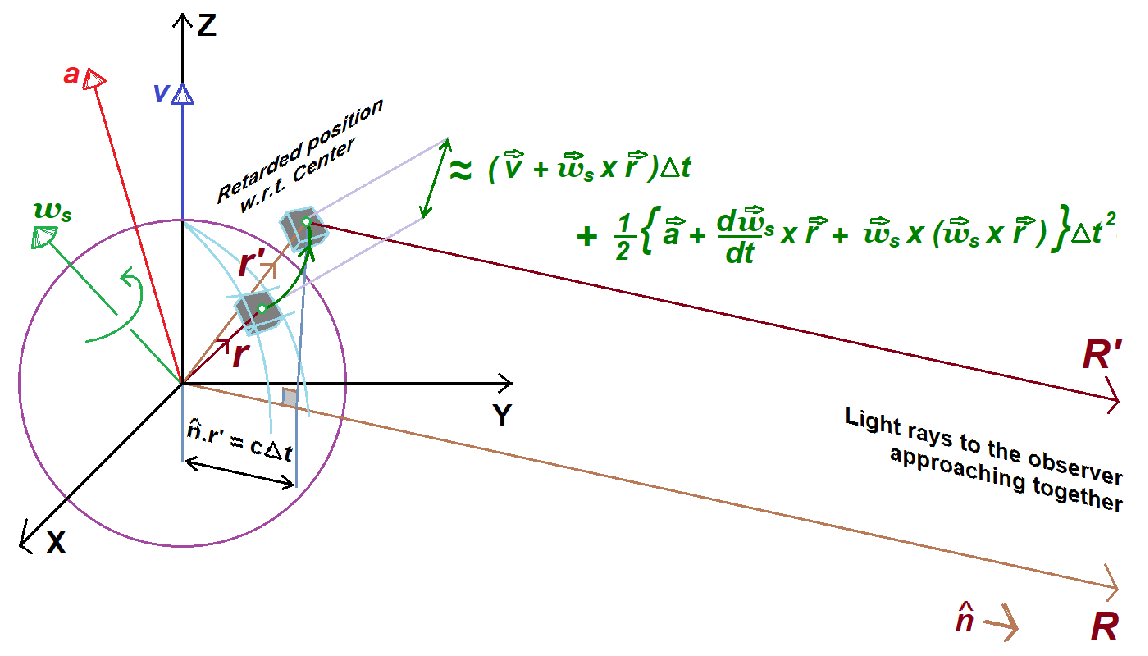}
\caption{(Color Online) Spinning point sphere in arbitrary motion.}
\label{fig:4}
\end{figure}

Due to linear and rotational velocities and accelerations, the volume-element would be subjected to retardation with reference to the center of the sphere. \bigskip

Thus, when a distant observer detects a light-ray from the center of the sphere, the ray that arrives from the volume-element appears to have originated from a slightly displaced position $ \textbf{r'} $. \bigskip

While, the amount of retardation satisfies the equation:

\begin{equation} \label{8}
c \Delta t \ = \boldsymbol{ \hat n } \cdot \textbf{r'}
\end{equation} \medskip

Where, $ \Delta t $ is the coordinate-time difference in which the volume-element moves from position $ \textbf{r} $ to $ \textbf{r'} $; and $ \boldsymbol{ \hat n } $ is the unit-vector pointing in the direction of faraway observer. \bigskip

If $ \boldsymbol{ v_r } $ and $ \boldsymbol{ a_r } $ being the net velocity and acceleration of the volume-element at position $ \textbf{r} $, respectively; the retarded-position $ \textbf{r'} $ and velocity at retarded-position $ \boldsymbol{ v_{r'} } $, are evaluated using the Taylor series as:

\begin{equation} \label{9}
\textbf{r'} \ = \textbf{r} \ + \boldsymbol{ v_r } \Delta t \ +  \dfrac{ 1 }{ 2 } \boldsymbol{ a_r } \Delta t^2 \ + \dfrac{ 1 }{ 6 } \boldsymbol{ \dot a_r } \Delta t^3 \ + \cdots \cdots
\end{equation} \medskip

and,

\begin{equation} \label{10}
\boldsymbol{ v_{r'} } \ = \boldsymbol{ v_r } \ + \boldsymbol{ a_r } \Delta t \ + \dfrac{ 1 }{ 2 } \boldsymbol{ \dot a_r } \Delta t^2 \ +  \dfrac{ 1 }{ 6 } \boldsymbol{ \ddot a_r } \Delta t^3 \ + \cdots \cdots
\end{equation} \medskip

Where, $ \boldsymbol{ \dot a_r } \ = d \boldsymbol{ a_r } / d t_{ret} $ and $ \boldsymbol{ \ddot a_r } \ = d^2 \boldsymbol{ a_r } / d t_{ret}^2 $ being the acceleration and jerk of the volume-element, respectively, evaluated at retarded time $ t_{ret} $. \bigskip

Further, taking spin into account, $ \boldsymbol{ v_r } $ would be the sum of the center's velocity and the rotational velocity of the volume-element with respect to the center, as given by:

\begin{equation} \label{11}
\boldsymbol{ v_r } \ = \textbf{v} \ + \boldsymbol{ \omega_s } \times \textbf{r}
\end{equation} \medskip

From, $ \boldsymbol{ v_r } $ above, $ \boldsymbol{ a_r } $ may be evaluated as:

\begin{equation} \label{12}
\begin{matrix}
\boldsymbol{ a_r } & \ = \dfrac{ d \boldsymbol{ v_r } }{ dt_{ret} } & \ = &  \textbf{a} \ + \dfrac{ d \boldsymbol{ \omega_s } }{ dt_{ret} } \times \textbf{r} \ + \boldsymbol{ \omega_s } \times ( \boldsymbol{ \omega_s } \times \textbf{r} ) \\
{}& {}& {}& {} \\
{}& {}& \ = &
\textbf{a} \ + \boldsymbol{ \dot \omega_s } \times \textbf{r} \ +  \left\{ ( \boldsymbol{ \omega_s } \cdot \textbf{r} ) \boldsymbol{ \omega_s } \ - { \omega_s }^2 \textbf{r} \right\} 
\end{matrix} 
\end{equation} \medskip

The first term in equation (\ref{12}) is instantaneous linear acceleration of the center of the sphere. The second term is the acceleration component caused by angular-acceleration. The third-term represents the centrifugal-acceleration. \bigskip

Furthermore, the jerk on the volume-element at position $ \textbf{r} $ can be evaluated as:

\begin{equation} \label{13}
\begin{matrix}
\boldsymbol{ \dot a_r } & \ = & \dfrac{ d \boldsymbol{ a_r } }{ dt_{ret} } \\
{}& {}& {} \\
{} & \ = & \dfrac{ d \textbf{a} }{ dt_{ret} } \ + \dfrac{ d^2 \boldsymbol{ \omega_s } }{ dt_{ret} ^2 } \times \textbf{r} \ + 2 \dfrac{ d \boldsymbol{ \omega_s } }{ dt_{ret} } \times \left( \boldsymbol{ \omega_s } \times \textbf{r} \right) \ + \boldsymbol{ \omega_s } \times \left( \dfrac{ d \boldsymbol{ \omega_s } }{ dt_{ret} } \times \textbf{r} \right) \\ 
{}& {}& \ + \boldsymbol{ \omega_s } \times \left\{ \boldsymbol{ \omega_s } \times ( \boldsymbol{ \omega_s } \times \textbf{r} ) \right\} \\
{}& {}& {} \\
{}& \ = & \boldsymbol{ \dot a } \ + \boldsymbol{ \ddot \omega_s } \times \textbf{r} \ + 2 \boldsymbol{ \dot \omega_s } \times \left( \boldsymbol{ \omega_s } \times \textbf{r} \right) \ + \boldsymbol{ \omega_s } \times \left( \boldsymbol{ \dot \omega_s } \times \textbf{r} \right) \\
{}& {}& \ - { \omega_s }^2 ( \boldsymbol{ \omega_s } \times \textbf{r} )
\end{matrix} 
\end{equation} \medskip

If $ \textbf{R} $ is the vector distance of the distant observer from the center of the point-charge, the vector distance of the observer from the retarded position of the volume-element is given by:

\begin{equation} \label{14}
\textbf{R'} \ = \textbf{R} \ - \textbf{r'}
\end{equation} \medskip

and, scalar-distance $ R' $ is given by:

\begin{equation} \label{15}
R' \ = \sqrt ( R^2 \ + r'^2 \ - 2 \textbf{R} \cdot \textbf{r'} ) \ = R \left\{ 1 \ + \dfrac{ r'^2 }{ R^2 } \ - \dfrac{ 2 }{ R } ( \boldsymbol{ \hat n } \cdot \textbf{r'} ) \right\} ^{ \dfrac{ 1 }{ 2 } }
\end{equation} \medskip

Equations (\ref{14}) and (\ref{15}) along with (\ref{9}) show that due to retardation, the sphere appears deformed to the distant observer. \bigskip

Moreover, the amount of apparent-deformation varies in complex manner depending on magnitudes of velocity, acceleration, angular-velocity of the spin, their relative orientations with reference to the observer, and their mutual interaction. \bigskip

In the following section we evaluate LW potentials for our spinning small rigid sphere in non-relativistic motion. We shall also demonstrate that in the point-particle limit $ r_q \rightarrow 0 $ if spin-angular-momentum were considered a bulk property of the point-charge, independent of its dimensions, the terms containing `classical-spin' survive and show up in the expression of LW potentials. \medskip

\section{LW 4-Potentials due to Spinning Point-Charge in motion:} \label{derive}

Using equation (\ref{5}) for apparently deformed dimensions and with $ \rho $ being the charge-density, the retarded 4-potentials due to the volume-element is given by:

\begin{equation} \label{16}
\begin{matrix}
d A^\mu ( R^\mu) & \ = &
\dfrac{ \rho u^\mu ( \textbf{r'} ) }{ 4 \pi \epsilon_0 c^2 } \dfrac{ r^2 d \Omega  d r }{ \left\{ - R'_\mu u^\mu ( \textbf{r'} ) \right\} } \\
{}& {}& {}\\
{}& \ = &
\dfrac{ \rho }{ 4 \pi \epsilon_0 c^2 } \dfrac{ \left[ c; \ \boldsymbol{ v_{r'} } \right] r^2 d \Omega  d r }{ \left\{ R' c \ - \textbf{R'} \cdot \boldsymbol{ v_{r'} } \right\} } 
\end{matrix} 
\end{equation} \medskip

Integrating equation (\ref{16}) over the entire volume of the sphere and applying the point-particle limit $ r_q \rightarrow 0 $ would yield LW potentials due to ``spinning point-charge in arbitrary motion''. \bigskip

However, the 4-vector terms $ u^\mu ( \textbf{r'} ) / \left\{ - R'_\mu u^\mu ( \textbf{r'} ) \right\} $ for the volume-element, which represent the apparent deformed geometry, need to be evaluated first. \bigskip

From equation (\ref{8}), $ \Delta t $ seems to be of the order of $ r' / c $. In the non-relativistic limit, i.e. for 'slow enough' velocity and acceleration ( $ v << c $, $ a \Delta t << c $ ), $ r' $ approaches $ r $. \bigskip

Further, $ r $ being the radial distance from the center of point-particle, is too small as compared to the distance from the observer (i.e. $ r << R $). Thus, in point-particle limit $ r_q \rightarrow 0 $, $ \Delta t $ would also be negligibly small; and $ \Delta t \approx r/ c $. \bigskip

However, as the order of angular-momentum goes as $ r^2 $, to examine spin dependence of LW potentials, we must include terms of at least the order $ \Delta t^2 $ in our approximation. \bigskip

Thus, equation (\ref{9}) for the apparent retardation and equation (\ref{10}) for the apparent velocity at retarded position, as sensed by the faraway observer, are approximated to the second powers of  $ \Delta t $, as:

\begin{equation} \label{17}
\textbf{r'} \ \approx \textbf{r} \ + \boldsymbol{ v_r } \Delta t \ + \dfrac{ 1 }{ 2 } \boldsymbol{ a_r } \Delta t^2
\end{equation} \medskip

and, 

\begin{equation} \label{18}
\boldsymbol{ v_{r'} } \ \approx \boldsymbol{ v_r } \ + \boldsymbol{ a_r } \Delta t \ + \dfrac{ 1 }{ 2 }  \boldsymbol{ \dot a_r } \Delta t^2
\end{equation} \medskip

From (\ref{14}) and (\ref{15}), along with (\ref{17}) and (\ref{18}), expanding the square-root term of (\ref{15}) by binomial theorem, and keeping only up to $ r^2 $ like terms (i.e. up to $ r^2 $, $ r \Delta t $ and $ \Delta t^2 $ ) , the quantity in denominator of (\ref{16}) $ - R'_\mu u^\mu ( \textbf{r'} ) $ may be evaluated as: 

\begin{equation} \label{19}
\begin{matrix}
- R'_\mu u^\mu ( \textbf{r'} ) & \ = & R' c \ - \textbf{R'} \cdot \boldsymbol{ v_{r'} } \\
{}& {}& {}\\
{}& \ = &  R  \left[\
c \ - \boldsymbol{ \hat n } \cdot \boldsymbol{ v_r } \ - \left( \boldsymbol{ \hat n } \cdot \boldsymbol{ a_r } \right) \Delta t \ - \dfrac{ 1 }{ 2 } \left( \boldsymbol{ \hat n } \cdot \boldsymbol{ \dot a_r } \right) \Delta t^2 \right] \\
{}& {}& {}\\
{}& {}& \ + \left[\
\left( \textbf{r} \cdot \boldsymbol{ v_r } \right) \ - c ( \boldsymbol{ \hat n } \cdot \textbf{r} ) \ - c \left( \boldsymbol{ \hat n } \cdot \boldsymbol{ v_r } \right) \Delta t \ + { v_{ \textbf{r} } }^2 \Delta t 
\right. \\
{}& {}& \left.
\ + \left( \textbf{r} \cdot \boldsymbol{ a_r } \right) \Delta t \ - \dfrac{ c }{ 2 } \left( \boldsymbol{ \hat n } \cdot \boldsymbol{ a_r } \right) \ + \dfrac{ 3 }{ 2 } \left( \boldsymbol{ v_r } \cdot \boldsymbol{ a_r } \right) 
\right] \\
{}& {}& {}\\
{}& {}& \ + \dfrac{ c }{ R } \left[\
\dfrac{ r^2 }{ 2 } \ - \left( \textbf{r} \cdot \boldsymbol{ v_r } \right) \Delta t \ + \dfrac{ 1 }{ 2 } { v_{ \textbf{r} } }^ 2 \Delta t^2 
\right. \\
{}& {}& \left.
\ - \dfrac{ 1 }{ 2 } \left( \boldsymbol{ \hat n } \cdot \boldsymbol{ v_r } \right)^2 \Delta t^2 \ - \dfrac{ 1 }{ 2 } ( \boldsymbol{ \hat n } \cdot \textbf{r} )^2 \ - \left( \boldsymbol{ \hat n } \cdot \boldsymbol{ v_r } \right) ( \boldsymbol{ \hat n } \cdot \textbf{r} ) \Delta t  
\right]
\end{matrix}
\end{equation} \medskip

Meanwhile, $ \Delta t $ may be evaluated by using equation (\ref{17}) with (\ref{8}), which yields the  quadratic equation in $ \Delta t $:

\begin{equation} \label{20}
\dfrac{ 1 }{ 2 } \left( \boldsymbol{ \hat n } \cdot \boldsymbol
{ a_r } \right) \Delta t^2 \ -   \left( c \ - \boldsymbol{ \hat n } \cdot \boldsymbol{ v_r } \right) \Delta t \ + \boldsymbol{ \hat n } \cdot \textbf{r} \ = 0
\end{equation} \medskip

solution of which comes out to be:

\begin{equation} \label{21}
\Delta t \ = \dfrac{ \left( c \ - \boldsymbol{ \hat n } \cdot \boldsymbol{ v_r } \right) }{ \boldsymbol{ \hat n } \cdot \boldsymbol{ a_r } }
\ \pm
\dfrac{ 1 }{ \boldsymbol{ \hat n } \cdot \boldsymbol{ a_r } } 
\sqrt { \left[ 
\left( c \ - \boldsymbol{ \hat n } \cdot \boldsymbol{ v_r } \right)^2
\ - 2 \left( \boldsymbol{ \hat n } \cdot \boldsymbol{ a_r } \right) ( \boldsymbol{ \hat n } \cdot \textbf{r} )
\right] }
\end{equation} \medskip

As $ \Delta t $ must vanish at $ r \ = 0 $, solving with negative sign and expanding the square-root with binomial theorem, we get the expression for $ \Delta t $, approximated up to $ r^2 $ terms as:

\begin{equation} \label{22}
\Delta t \ \approx \dfrac{ ( \boldsymbol{ \hat n } \cdot \textbf{r} ) }{ \left( c \ -  \boldsymbol{ \hat n } \cdot \boldsymbol{ v_r } \right) } 
\ +
\dfrac{ 1 }{ 2 } \dfrac{ ( \boldsymbol{ \hat n } \cdot \textbf{r} )^2 \left( \boldsymbol{ \hat n } \cdot \boldsymbol
{ a_r } \right) }{ \left( c \ - \boldsymbol{ \hat n } \cdot \boldsymbol{ v_r } \right)^3 }
\end{equation} \medskip

Using expression of $ \Delta t $ of equation (\ref{22}) in equation (\ref{18}), velocity of the volume-element at retarded position $ \textbf{r'} $, approximated up to $ r^2 $ terms, comes out to be:

\begin{equation} \label{23}
\begin{matrix}
\boldsymbol{ v_{r'} } \ \approx \boldsymbol{ v_r } \ + \dfrac{ ( \boldsymbol{ \hat n } \cdot \textbf{r} ) \boldsymbol{ a_r } }{ \left( c \ - \boldsymbol{ \hat n } \cdot \boldsymbol{ v_r } \right) }
& \ + \dfrac{ 1 }{ 2 } \dfrac{ ( \boldsymbol{ \hat n } \cdot \textbf{r} )^2 \left( \boldsymbol{ \hat n } \cdot \boldsymbol{ a_r } \right) \boldsymbol{ a_r } }{ \left( c \ - \boldsymbol{ \hat n } \cdot \boldsymbol{ v_r } \right)^ 3 } \\
{} & {} \\
{} & \ + \dfrac{ 1 }{ 2 } \dfrac{ ( \boldsymbol{ \hat n } \cdot \textbf{r} )^2 \boldsymbol{ \dot a_r } }{ \left( c \ - \boldsymbol{ \hat n } \cdot \boldsymbol{ v_r } \right)^2 }
\end{matrix}
\end{equation} \medskip

Using equations (\ref{11}), (\ref{12}) and (\ref{13}) in (\ref{23}) and up to the $ r^2 $ terms, we get:

\begin{equation} \label{24}
\begin{matrix}
\boldsymbol{ v_{r'} } & \ \approx\
\textbf{v} \ + \boldsymbol{ \omega_s } \times \textbf{r}
\ + \dfrac{ ( \boldsymbol{ \hat n } \cdot \textbf{r} ) \textbf{a} }{ ( c \ - \boldsymbol{ \hat n } \cdot \textbf{v} ) } 
\ + \dfrac{ ( \boldsymbol{ \hat n } \cdot \textbf{r} ) ( \textbf{r} \cdot \langle \boldsymbol{ \hat n } \times \boldsymbol{ \omega_s } \rangle ) \textbf{a} }{ ( c \ - \boldsymbol{ \hat n } \cdot \textbf{v} )^2 } \\
{}& {}\\
{} & 
\ - \dfrac{ ( \boldsymbol{ \hat n } \cdot \textbf{r} ) ( \textbf{r} \times \boldsymbol{ \dot \omega_s } ) }{ ( c \ - \boldsymbol{ \hat n } \cdot \textbf{v} ) } 
\ + \dfrac{ ( \boldsymbol{ \hat n } \cdot \textbf{r} ) ( \boldsymbol{ \omega_s } \cdot \textbf{r} ) \boldsymbol{ \omega_s } }{ ( c \ - \boldsymbol{ \hat n } \cdot \textbf{v} ) }
\ - \dfrac{ ( \boldsymbol{ \hat n } \cdot \textbf{r} ) { \omega_s }^2 \textbf{r} }{ ( c \ - \boldsymbol{ \hat n } \cdot \textbf{v} ) } \\
{}& {}\\
{} & 
\ + \dfrac{ 1 }{ 2 } \dfrac{ ( \boldsymbol{ \hat n } \cdot \textbf{r} )^2 \left\{ \boldsymbol{ \hat n } \cdot \textbf{a} \right\} \textbf{a} }{ ( c \ - \boldsymbol{ \hat n } \cdot \textbf{v} )^3 } 
\ + \dfrac{ 1 }{ 2 } \dfrac{ ( \boldsymbol{ \hat n } \cdot \textbf{r} )^2 \boldsymbol{ \dot a } }{ ( c \ - \boldsymbol{ \hat n } \cdot \textbf{v} )^2 }
\end{matrix}
\end{equation} \medskip

Further, using expression of $ \Delta t $ from equation (\ref{22}), and inverting equation (\ref{19}) with binomial-expansion in fractions of $ \left( c \ - \boldsymbol{ \hat n } \cdot \boldsymbol{ v_r } \right) $, up to $ r^2 $ like terms; we get:

\begin{equation} \label{25}
\begin{matrix}
\dfrac{ 1 }{ \left\{ - R'_\mu u^\mu ( \textbf{r'} ) \right\} } & \ \approx \dfrac{ 1 }{ R } \left[ \ 
\dfrac{ 1 }{ \left( c \ - \boldsymbol{ \hat n } \cdot \boldsymbol{ v_r } \right) }
\ + \dfrac{ \left( \boldsymbol{ \hat n } \cdot \boldsymbol{ a_r } \right) ( \boldsymbol{ \hat n } \cdot \textbf{r} ) }{ \left( c \ - \boldsymbol{ \hat n } \cdot \boldsymbol{ v_r } \right)^3 } 
\ + \dfrac{ 3 }{ 2 } \dfrac{ \left( \boldsymbol{ \hat n } \cdot \boldsymbol{ a_r } \right)^2 (\boldsymbol{ \hat n } \cdot \textbf{r} )^2 }{ \left( c \ - \boldsymbol{ \hat n } \cdot \boldsymbol{ v_r } \right)^5 } 
\right.
\\
{} & \left.
\ + \dfrac{ 1 }{ 2 } \dfrac{ \left( \boldsymbol{ \hat n } \cdot \boldsymbol{ a_r } \right) ( \boldsymbol{ \hat n } \cdot \textbf{r} )^2 }{ \left( c \ - \boldsymbol{ \hat n } \cdot \boldsymbol{ v_r } \right)^4 }
\right] \\
{} & {} \\
{} &
\ + \dfrac{ 1 }{ R^2 } \left[\
\dfrac{ c^2 ( \boldsymbol{ \hat n } \cdot \textbf{r} ) }{ \left( c \ - \boldsymbol{ \hat n } \cdot \boldsymbol{ v_r } \right)^3 }
\ - \dfrac{ \left( \textbf{r} \cdot \boldsymbol{ v_r } \right) }{ \left( c \ - \boldsymbol{ \hat n } \cdot \boldsymbol{ v_r } \right)^2 } 
\ + \dfrac{ 5 c^2 }{ 2 } \dfrac{ \left( \boldsymbol{ \hat n } \cdot \boldsymbol{ a_r } \right) (\boldsymbol{ \hat n } \cdot \textbf{r} )^2 }{ \left( c \ - \boldsymbol{ \hat n } \cdot \boldsymbol{ v_r } \right)^5 }
\right.
\\
{} & \left.
\ - \dfrac{ { v_{ \textbf{r} } }^2 ( \boldsymbol{ \hat n } \cdot \textbf{r} ) }{ \left( c \ - \boldsymbol{ \hat n } \cdot \boldsymbol{ v_r } \right)^3 }
\ - \dfrac{ 2 \left( \boldsymbol{ \hat n } \cdot \boldsymbol{ a_r } \right) \left( \textbf{r} \cdot \boldsymbol{ v_r } \right) ( \boldsymbol{ \hat n } \cdot \textbf{r} ) }{ \left( c \ - \boldsymbol{ \hat n } \cdot \boldsymbol{ v_r } \right)^3 } 
\ - \dfrac{ \left( \textbf{r} \cdot \boldsymbol{ a_r } \right) ( \boldsymbol{ \hat n } \cdot \textbf{r} ) }{ \left( c \ - \boldsymbol{ \hat n } \cdot \boldsymbol{ v_r } \right)^3 }
\right.
\\
{} & \left.
\ - \dfrac{ 5 }{ 2 } \dfrac{ { v_{\textbf{r}} }^2 \left( \boldsymbol{ \hat n } \cdot \boldsymbol{ a_r } \right) ( \boldsymbol{ \hat n } \cdot \textbf{r} )^2 }{ \left( c \ - \boldsymbol{ \hat n } \cdot \boldsymbol{ v_r } \right)^5 } 
\ - \dfrac{ 3 }{ 2 } \dfrac{ \left( \boldsymbol{ v_r } \cdot \boldsymbol{ a_r } \right) (\boldsymbol{ \hat n } \cdot \textbf{r} )^2 }{ \left( c \ - \boldsymbol{ \hat n } \cdot \boldsymbol{ v_r } \right)^4 }
\right] \\
{} & {} \\
{} &
\ + \dfrac{ c }{ R^3 } \left[\  
\dfrac{ \left( \textbf{r} \cdot \boldsymbol{ v_r } \right) ( \boldsymbol{ \hat n } \cdot \textbf{r} ) }{ \left( c \ - \boldsymbol{ \hat n } \cdot \boldsymbol{ v_r } \right)^3 }
\ - \dfrac{ 1 }{ 2 } \dfrac{ r^2 }{ \left( c \ - \boldsymbol{ \hat n } \cdot \boldsymbol{ v_r } \right)^2 }
\ - \dfrac{ 1 }{ 2 } \dfrac{ \left( \boldsymbol{ v_r } \right)^2 ( \boldsymbol{ \hat n } \cdot \textbf{r} )^2 }{ \left( c \ - \boldsymbol{ \hat n } \cdot \boldsymbol{ v_r } \right)^4 }
\right.
\\
{} & \left.
\ + \dfrac{ c^2 }{ 2 } \dfrac{ ( \boldsymbol{ \hat n } \cdot \textbf{r} )^2 }{ \left( c \ - \boldsymbol{ \hat n } \cdot \boldsymbol{ v_r } \right)^4 }
\ - \dfrac{ 2 c \left( \textbf{r} \cdot \boldsymbol{ v_r } \right) ( \boldsymbol{ \hat n } \cdot \textbf{r} ) }{ \left( c \ - \boldsymbol{ \hat n } \cdot \boldsymbol{ v_r } \right)^4 }
\ + \dfrac{ 2 c \left( \boldsymbol{ \hat n } \cdot \boldsymbol{ v_r } \right) ( \boldsymbol{ \hat n } \cdot \textbf{r} )^2 }{ \left( c \ - \boldsymbol{ \hat n } \cdot \boldsymbol{ v_r } \right)^4 }
\right.
\\
{} & \left.
\ - \dfrac{ 2 c { v_{ \textbf{r} } }^2 ( \boldsymbol{ \hat n } \cdot \textbf{r} )^2 }{ \left( c \ - \boldsymbol{ \hat n } \cdot \boldsymbol{ v_r } \right)^5 }
\ - \dfrac{ 2 { v_{ \textbf{r} } }^2 \left( \textbf{r} \cdot \boldsymbol{ v_r } \right) ( \boldsymbol{ \hat n } \cdot \textbf{r} ) }{ \left( c \ - \boldsymbol{ \hat n } \cdot \boldsymbol{ v_r } \right)^4 }
\right]
\end{matrix}
\end{equation} \medskip

Now, inserting spin dependent quantities by using equations (\ref{11}), (\ref{12}) and (\ref{13}), in (\ref{25}), and carrying out binomial-expansion of the denominator in fractions of $ \left( c \ - \boldsymbol{ \hat n } \cdot \textbf{v} \right) $, and keeping the terms up to $ r^2 $, we get:

\begin{equation} \label{26}
\begin{matrix}
\dfrac{ 1 }{ \left\{ - R'_\mu u^\mu ( \textbf{r'} ) \right\} } \ \approx
\dfrac{ 1 }{ R } \left[
\dfrac{ 1 }{ ( c \ - \boldsymbol{ \hat n } \cdot \textbf{v} ) } 
\left\{ 1 
\ + \dfrac{ \boldsymbol{ \hat n } \cdot \langle \textbf{r} \times \boldsymbol{ \omega_s } \rangle }{ ( c \ - \boldsymbol{ \hat n } \cdot \textbf{v} ) } 
\ + \dfrac{ ( \boldsymbol{ \hat n } \cdot \langle \textbf{r} \times \boldsymbol{ \omega_s } \rangle )^2 }{ ( c \ - \boldsymbol{ \hat n } \cdot \textbf{v} )^2 }
\right\}
\right.
\\ 
\left.
\ + \dfrac{ ( \boldsymbol{ \hat n } \cdot \textbf{r} ) }{ ( c \ - \boldsymbol{ \hat n } \cdot \textbf{v} )^3 }
\left\{ 
\boldsymbol{ \hat n } \cdot \left( 
\textbf{a} 
\ + { \boldsymbol{ \dot \omega_s } \times \textbf{r} }  
\ + ( \boldsymbol{ \omega_s } \cdot \textbf{r} ) \boldsymbol{ \omega_s }
\ - { \omega_s }^2 \textbf{r} 
\right) \right\}
\left\{ 1 
\ + \dfrac{ 3 ( \boldsymbol{ \hat n } \cdot \langle \boldsymbol{ \omega_s } \times \textbf{r} \rangle ) }{ ( c \ - \boldsymbol{ \hat n } \cdot \textbf{v} ) }
\right\}
\right.
{} \\ 
\left.
\ + \dfrac{ 3 }{ 2 } \dfrac{ ( \boldsymbol{ \hat n } \cdot \textbf{a} )^2 ( \boldsymbol{ \hat n } \cdot \textbf{r} )^2 }{ ( c \ - \boldsymbol{ \hat n } \cdot \textbf{v} )^5 } 
\ + \dfrac{ 1 }{ 2 } \dfrac{ ( \boldsymbol{ \hat n } \cdot \textbf{a} ) ( \boldsymbol{ \hat n } \cdot \textbf{r} )^2 }{ ( c \ - \boldsymbol{ \hat n } \cdot \textbf{v} )^4 }
\right]
{}\\
{}\\
\ + \dfrac{ 1 }{ R^2 } \left[
{ c^2 ( \boldsymbol{ \hat n } \cdot \textbf{r} ) }{ ( c \ - \boldsymbol{ \hat n } \cdot \textbf{v} )^3 } \left\{ 1 
\ + \dfrac{ 3 ( \boldsymbol{ \hat n } \cdot \langle \boldsymbol{ \omega_s } \times \textbf{r} \rangle ) }{ ( c \ - \boldsymbol{ \hat n } \cdot \textbf{v} ) }
\right\}
\right.
{} \\
\left.
\ - \dfrac{ 1 }{ ( c \ - \boldsymbol{ \hat n } \cdot \textbf{v} )^2 } \left\{
\textbf{r} \cdot ( \textbf{v} \ + \boldsymbol{ \omega_s } \times \textbf{r} ) 
\right\}
\left\{ 1 
\ + \dfrac{ 2 ( \boldsymbol{ \hat n } \cdot \langle \boldsymbol{ \omega_s } \times \textbf{r} \rangle ) }{ ( c \ - \boldsymbol{ \hat n } \cdot \textbf{v} ) }
\right\}
\right.
{} \\
\left.
\ + \dfrac{ 5 c^2 }{ 2 } \dfrac{ ( \boldsymbol{ \hat n } \cdot \textbf{a} ) ( \boldsymbol{ \hat n } \cdot \textbf{r} )^2 }{ ( c \ - \boldsymbol{ \hat n } \cdot \textbf{v}  )^5 } 
\ - \dfrac{ ( \boldsymbol{ \hat n } \cdot \textbf{r} ) }{ ( c \ - \boldsymbol{ \hat n } \cdot \textbf{v} )^3 } \left\{ 
\textbf{v}  \ + \boldsymbol{ \omega_s } \times \textbf{r}  \right\}^2 \left\{ 1 
\ + \dfrac{ 3 ( \boldsymbol{ \hat n } \cdot \langle \boldsymbol{ \omega_s } \times \textbf{r} \rangle ) }{ ( c \ - \boldsymbol{ \hat n } \cdot \textbf{v} ) }
\right\}
\right.
{} \\
\left.
\ - \dfrac{ 2 ( \boldsymbol{ \hat n } \cdot \textbf{a} ) ( \textbf{r} \cdot \textbf{v} ) (\boldsymbol{ \hat n } \cdot \textbf{r} ) }{ ( c \ - \boldsymbol{ \hat n } \cdot \textbf{v} )^3 }  
\ - \dfrac{ ( \textbf{r} \cdot \textbf{a} ) ( \boldsymbol{ \hat n } \cdot \textbf{r} ) }{ ( c \ - \boldsymbol{ \hat n } \cdot \textbf{v} )^3 }
\right.
{} \\
\left.
\ - \dfrac{ 5 }{ 2 } \dfrac{ v^2 ( \boldsymbol{ \hat n } \cdot \textbf{a} ) ( \boldsymbol{ \hat n } \cdot \textbf{r} )^2 }{ ( c \ - \boldsymbol{ \hat n } \cdot \textbf{v} )^5 } 
\ - \dfrac{ 3 }{ 2 } \dfrac{ ( \textbf{v} \cdot \textbf{a} ) ( \boldsymbol{ \hat n } \cdot \textbf{r} )^2 }{ ( c \ - \boldsymbol{ \hat n } \cdot \textbf{v} )^4 }
\right]
{}\\
{}\\
\ + \dfrac{ c }{ R^3 } \left[ 
\dfrac{ ( \textbf{r} \cdot \textbf{v} ) ( \boldsymbol{ \hat n } \cdot \textbf{r} ) }{ ( c \ - \boldsymbol{ \hat n } \cdot \textbf{v} )^3 } 
\ - \dfrac{ 1 }{ 2 } \dfrac{ r^2 }{ ( c \ - \boldsymbol{ \hat n } \cdot \textbf{v} )^2 } 
\ - \dfrac{ 1 }{ 2 } \dfrac{ v^2 ( \boldsymbol{ \hat n } \cdot \textbf{r} )^2 }{ ( c \ - \boldsymbol{ \hat n } \cdot \textbf{v} )^4 } 
\ + \dfrac{ c ^2 }{ 2 } \dfrac{ ( \boldsymbol{ \hat n } \cdot \textbf{r} )^2 }{ ( c \ - \boldsymbol{ \hat n } \cdot \textbf{v} )^4 } 
\right.
{}\\
\left.
\ - \dfrac{ 2 c ( \textbf{r} \cdot \textbf{v} ) ( \boldsymbol{ \hat n } \cdot \textbf{r} ) }{ ( c \ - \boldsymbol{ \hat n } \cdot \textbf{v} )^4 } 
\ + \dfrac{ 2 c ( \boldsymbol{ \hat n } \cdot \textbf{v} ) ( \boldsymbol{ \hat n } \cdot \textbf{r} )^2 }{ ( c \ - \boldsymbol{ \hat n } \cdot \textbf{v} )^4 }
\ - \dfrac{ 2 c v^2 ( \boldsymbol{ \hat n } \cdot \textbf{r} )^2 }{ ( c \ - \boldsymbol{ \hat n } \cdot \textbf{v} )^5 } 
\ - \dfrac{ 2 v^2 ( \textbf{r}  \cdot \textbf{v} ) ( \boldsymbol{ \hat n } \cdot \textbf{r} ) }{ ( c \ - \boldsymbol{ \hat n } \cdot \textbf{v} )^4 } 
\right]
\end{matrix}
\end{equation} \medskip

Now LW Potentials can be simply evaluated calculating  $ u^\mu ( \textbf{r'} ) / \left\{ - R'_\mu u^\mu ( \textbf{r'} ) \right\} $ using product of equations (\ref{26}) above, and (\ref{24}), plugging it into equation (\ref{16}), performing integration over the entire volume of the sphere, and lastly applying the point-particle limit $ r_q \rightarrow 0 $. \bigskip

Though, it appears that the resulting expression of 4-vector $ u^\mu ( \textbf{r'} ) / \left\{ - R'_\mu u^\mu ( \textbf{r'} ) \right\} $ such obtained, would be be fairly lengthy; many terms may be omitted by inspection, using formulae for known “spherical integrals”, and considering the terms in the context of the point-particle limit $ r_q \rightarrow 0 $. \bigskip

Following are the known spherical integrals:

\begin{equation} \label{27}
\left. 
\begin{matrix}
\iint_{ \Omega } ( \textbf{r} \cdot \textbf{A} ) d \Omega \ = 0; \\ \\
\iint_{ \Omega } ( \textbf{r} \times \textbf{A} ) d \Omega \ = 0; \\ \\
\iint_{ \Omega } ( \textbf{r}  \cdot \textbf{A} )( \textbf{r}  \cdot \textbf{B} ) d \Omega \ = \dfrac{ 4 \pi }{ 3 } r^2 ( \textbf{A} \cdot \textbf{B} ); \\ \\
\iint_{ \Omega } ( \textbf{r} \cdot \textbf{A} )( \textbf{r}  \times \textbf{B} ) d \Omega \ = \dfrac{ 4 \pi }{ 3 } r^2 ( \textbf{A} \times \textbf{B} ); \\ \\
\iint_{ \Omega } ( \textbf{r} \cdot \textbf{A} )\textbf{r} d \Omega \ = \dfrac{ 4 \pi }{ 3 } r^2 \textbf{A};
\end{matrix}
\ \right\}
\end{equation} \medskip

Equations (\ref{27}) show that the spherical-integrals of terms that contain single $ r $ vanish. Further, the terms that contain $ r $ but are without $ \omega_s $ would approach 0  in the limit $ r_q \rightarrow 0 $. \bigskip

Thus, by identifying and writing only non-vanishing terms from the product of equations (\ref{26}) and (\ref{24}), and inserting them into (\ref{16}), LW 4-potentials due to the spinning point-charge in arbitrary motion can be evaluated by:

\begin{equation} \label{28}
\begin{matrix}
A^\mu ( R^\mu ) \ = 
\lim_{r_q \to \ 0 } \dfrac{ \rho }{ 4 \pi \epsilon_0 c^2 } \int_{r} r^2 dr \iint_{ \Omega } \left[
\dfrac{ \left[ c; \ \textbf{v} \right] }{ R } \left\{
\dfrac{ 1 }{ ( c \ - \boldsymbol{ \hat n } \cdot \textbf{v} ) }
\right. \right.
{} \\
\left. \left.
\ + \dfrac{ ( \textbf{r} \cdot \langle \boldsymbol{ \hat n } \times \boldsymbol{ \omega_s } \rangle )^2 }{ ( c \ - \boldsymbol{ \hat n } \cdot \textbf{v} )^3 } 
\ + \dfrac{ ( \boldsymbol{ \hat n } \cdot \textbf{r} ) ( \textbf{r}  \cdot \langle \boldsymbol{ \hat n } \times \boldsymbol{ \dot \omega_s } \rangle ) }{ ( c \ - \boldsymbol{ \hat n } \cdot \textbf{v} )^3 } 
\ + \dfrac{ ( \boldsymbol{ \hat n } \cdot \boldsymbol{ \omega_s } ) ( \boldsymbol{ \hat n } \cdot \textbf{r} ) (\textbf{r} \cdot \boldsymbol{ \omega_s } ) }{ ( c \ - \boldsymbol{ \hat n } \cdot \textbf{v} )^3 }
\right. \right.
{} \\
\left. \left.
\ - \dfrac{ { \omega_s }^2 ( \boldsymbol{ \hat n } \cdot \textbf{r}  )^2 }{ ( c \ - \boldsymbol{ \hat n } \cdot \textbf{v} )^3 }
\ + \dfrac{ 3 ( \boldsymbol{ \hat n } \cdot \textbf{a} ) ( \boldsymbol{ \hat n } \cdot \textbf{r} ) ( \textbf{r} \cdot \langle \boldsymbol{ \hat n } \times \boldsymbol{ \omega_s } \rangle ) }{ ( c \ - \boldsymbol{ \hat n } \cdot \textbf{v} )^4 }
\right\}
\right.
{} \\
{} \\
\left.
\ + \dfrac{ \left[ c; \ \textbf{v} \right] }{ R^2 } \left\{
\dfrac{ 3 c^2 ( \boldsymbol{ \hat n } \cdot \textbf{r} ) ( \textbf{r} \cdot \langle \boldsymbol{ \hat n } \times \boldsymbol{ \omega_s } \rangle ) }{ ( c \ - \boldsymbol{ \hat n } \cdot \textbf{v} )^4 } 
\ - \dfrac{ 2 ( \textbf{r} \cdot \textbf{v} ) ( \textbf{r} \cdot \langle \boldsymbol{ \hat n } \times \boldsymbol{ \omega_s } \rangle ) }{ ( c \ - \boldsymbol{ \hat n } \cdot \textbf{v} )^3 } 
\right. \right.
{} \\
\left. \left.
\ - \dfrac{ 2 ( \boldsymbol{ \hat n } \cdot \textbf{r} ) ( \textbf{r} \cdot \langle \textbf{v} \times \boldsymbol{ \omega_s } \rangle ) }{ ( c \ - \boldsymbol{ \hat n } \cdot \textbf{v} )^3 } 
\ - \dfrac{ 3 v^2 ( \boldsymbol{ \hat n } \cdot \textbf{r} ) ( \textbf{r} \cdot \langle \boldsymbol{ \hat n } \times \boldsymbol{ \omega_s } \rangle ) }{ ( c \ - \boldsymbol{ \hat n } \cdot \textbf{v} )^4 } 
\right\}
\right.
{} \\
{} \\
\left.
\ - \dfrac{ \left[ 0; \ \textbf{r} \times \boldsymbol{ \omega_s } \right] }{ R } \left\{
\dfrac{ (\boldsymbol{ \hat n } \cdot \textbf{r} ) }{ ( c \ - \boldsymbol{ \hat n } \cdot \textbf{v} )^2 }
\ + \dfrac{ ( \textbf{r} \cdot \langle \boldsymbol{ \hat n } \times \boldsymbol{ \omega_s } \rangle ) }{ ( c \ - \boldsymbol{ \hat n } \cdot \textbf{v} )^2 } 
\ + \dfrac{ ( \boldsymbol{ \hat n } \cdot \textbf{a} ) ( \boldsymbol{ \hat n } \cdot \textbf{r} ) }{ ( c \ - \boldsymbol{ \hat n } \cdot \textbf{v} )^3 }
\right\}
\right.
{} \\
{} \\
\left.
\ + \dfrac{ \left[ 0; \ \textbf{r} \times \boldsymbol{ \omega_s } \right] }{ R^2 } \left\{
\ - \dfrac{ c^2 (\boldsymbol{ \hat n } \cdot \textbf{r} ) }{ ( c \ - \boldsymbol{ \hat n } \cdot \textbf{v} )^3 }
\ +  \dfrac{ ( \textbf{r} \cdot \textbf{v} ) }{ ( c \ - \boldsymbol{ \hat n } \cdot \textbf{v} )^2 } 
\ + \dfrac{ v^2 ( \boldsymbol{ \hat n } \cdot \textbf{r} ) }{ ( c \ - \boldsymbol{ \hat n } \cdot \textbf{v} )^3 }
\right\}
\right.
{} \\
{} \\
\left.
\ + \dfrac{ \left[ 0; \ \textbf{a}  \right] }{ R } \left\{
\dfrac{ (\boldsymbol{ \hat n } \cdot \textbf{r} ) ( \textbf{r}  \cdot \langle \boldsymbol{ \hat n } \times \boldsymbol{ \omega_s } \rangle ) }{ ( c \ - \boldsymbol{ \hat n } \cdot \textbf{v} )^3 }
\ + \dfrac{ ( \boldsymbol{ \hat n } \cdot \textbf{r} ) ( \textbf{r}  \cdot \langle \boldsymbol{ \hat n } \times \boldsymbol{ \omega_s } \rangle ) }{ ( c \ - \boldsymbol{ \hat n } \cdot \textbf{v} )^3 }
\right\}
\right.
{} \\
{} \\
\left.
\ + \dfrac{ 1 }{ R } \left\{
\dfrac{ ( \boldsymbol{ \hat n } \cdot \textbf{r} ) ( \textbf{r}  \cdot \boldsymbol{ \omega_s } ) }{ ( c \ - \boldsymbol{ \hat n } \cdot \textbf{v} )^2 } { \left[ 0; \ \boldsymbol{ \omega_s } \right] }
\ - \dfrac{ { \omega_s }^2 ( \boldsymbol{ \hat n } \cdot \textbf{r}  ) }{ ( c \ - \boldsymbol{ \hat n } \cdot \textbf{v} )^2 } { \left[ 0; \ \textbf{r} \right] }
\right\}
\right] d \Omega
\end{matrix}
\end{equation} \medskip

Here, we have used the triple product identities: $ \boldsymbol{ \hat n } \cdot \langle \boldsymbol{ \omega_s } \times \textbf{r} \rangle \ = \textbf{r} \cdot \langle \boldsymbol{ \hat n } \times \boldsymbol{ \omega_s } \rangle $ and $ \boldsymbol{ \omega_s } \times \langle \boldsymbol{ \omega_s } \times \textbf{r} \rangle \ = ( \textbf{r} \cdot \boldsymbol{ \omega_s } ) \boldsymbol{ \omega_s } \ - { \omega_s }^2 \textbf{r} $. \bigskip

Evaluating, spherical-integrals by using the set of equation (\ref{27}) first, we get:

\begin{equation} \label{29}
\begin{matrix}
A^\mu ( R^\mu ) \ =  \lim_{r_q \to \ 0 } \dfrac{ \rho }{ 4 \pi \epsilon_0 c^2 } \int_{r} r^2 dr \left[
\dfrac{ \left[ c; \ \textbf{v} \right] }{ R } \left\{
\dfrac{ 4 \pi }{ ( c \ - \boldsymbol{ \hat n } \cdot \textbf{v} ) }
\right. \right.
\\
\left. \left.
\ + \dfrac{ 4 \pi r^2 ( \boldsymbol{ \hat n } \times \boldsymbol{ \omega_s } )^2 }{ 3 ( c \ - \boldsymbol{ \hat n } \cdot \textbf{v} )^3 } 
\ + \dfrac{ 4 \pi r^2 ( \boldsymbol{ \hat n } \cdot \langle \boldsymbol{ \hat n } \times \boldsymbol{ \dot \omega_s } \rangle ) }{ 3 ( c \ - \boldsymbol{ \hat n } \cdot \textbf{v} )^3 }
\ + \dfrac{ 4 \pi r^2 ( \boldsymbol{ \hat n } \cdot \boldsymbol{ \omega_s } )^2 }{ 3 ( c \ - \boldsymbol{ \hat n } \cdot \textbf{v}  )^3 }
\right. \right.
\\
\left. \left.
\ - \dfrac{ 4 \pi r^2 { \omega_s }^2 }{ 3 ( c \ - \boldsymbol{ \hat n } \cdot \textbf{v} )^3 }
\ + \dfrac{ 4 \pi r^2  ( \boldsymbol{ \hat n } \cdot \textbf{a} ) ( \boldsymbol{ \hat n } \cdot \langle \boldsymbol{ \hat n } \times \boldsymbol{ \omega_s } \rangle ) }{ ( c \ - \boldsymbol{ \hat n } \cdot \textbf{v} )^4 } 
\right\}
\right.
\\
\\
\left.
\ + \dfrac{ \left[ c; \ \textbf{v}  \right] }{ R^2} \left\{
\dfrac{ 4 \pi r^2 c^2 (\boldsymbol{ \hat n } \cdot \langle \boldsymbol{ \hat n } \times \boldsymbol{ \omega_s } \rangle ) }{ ( c \ - \boldsymbol{ \hat n } \cdot \textbf{v} )^4 } 
\ - \dfrac{ 8 \pi r^2 ( \textbf{v} \cdot \langle \boldsymbol{ \hat n } \times \boldsymbol{ \omega_s } \rangle ) }{ 3 ( c \ - \boldsymbol{ \hat n } \cdot \textbf{v} )^3 } 
\right. \right.
\\
\left. \left.
\ - \dfrac{ 8 \pi r^2 ( \boldsymbol{ \hat n } \cdot \langle \textbf{v} \times \boldsymbol{ \omega_s } \rangle ) }{ 3 ( c \ - \boldsymbol{ \hat n } \cdot \textbf{v} )^3 } 
\ - \dfrac{ 4 \pi r^2 v^2 ( \boldsymbol{ \hat n } \cdot \langle \boldsymbol{ \hat n } \times \boldsymbol{ \omega_s } \rangle ) }{ ( c \ - \boldsymbol{ \hat n } \cdot \textbf{v} )^4 } 
\right\}
\right.
\\
\\
\left.
\ - \dfrac{ \left[ 0; \ \boldsymbol{ \hat n } \times \boldsymbol{ \omega_s } \right] }{ R } \left\{
\dfrac{ 4 \pi r^2 ( \boldsymbol{ \hat n } \cdot \textbf{a} ) }{ 3 ( c \ - \boldsymbol{ \hat n } \cdot \textbf{v} )^3 }
\right\}
\right.
\\
\\
\left.
\ + \dfrac{ \left[ 0; \ \boldsymbol{ \hat n } \times \boldsymbol{ \omega_s } \right] }{ R^2 } \left\{ 
\ - \dfrac{ 4 \pi r^2 c^2 }{ 3 ( c \ - \boldsymbol{ \hat n } \cdot \textbf{v} )^3 }
\ + \dfrac{ 4 \pi r^2 v^2 }{ 3 ( c \ - \boldsymbol{ \hat n } \cdot \textbf{v} )^3 }
\right\}
\right.
\\
\\
\left.
\ + \dfrac{ \left[ 0; \ \textbf{a} \right] }{ R } \left\{
\dfrac{ 4 \pi r^2 ( \boldsymbol{ \hat n } \cdot \langle \boldsymbol{ \hat n } \times \boldsymbol{ \omega_s } \rangle ) }{ 3 ( c \ - \boldsymbol{ \hat n } \cdot \textbf{v} )^3 } 
\ + \dfrac{ 4 \pi r^2 ( \boldsymbol{ \hat n } \cdot \langle \boldsymbol{ \hat n } \times \boldsymbol{ \omega_s } \rangle ) }{ 3 ( c \ - \boldsymbol{ \hat n } \cdot \textbf{v} )^3 } 
\right\}
\right.
\\
\\
\left.
\ - \dfrac{ \left[ 0; \ \boldsymbol{ \hat n } \times \boldsymbol{ \dot \omega_s } \right] }{ R } \left\{
\dfrac{ 4 \pi r^2 }{ 3 ( c \ - \boldsymbol{ \hat n } \cdot \textbf{v} )^2 } 
\right\}
\ + \dfrac{ \left[ 0; \ \boldsymbol{ \omega_s } \right] }{ R } \left\{
\dfrac{ 4 \pi r^2 ( \boldsymbol{ \hat n } \cdot \boldsymbol{ \omega_s } ) }{ 3 ( c \ - \boldsymbol{ \hat n } \cdot \textbf{v} )^2 } 
\right\}
\right.
\\
\\
\left.
\ - \dfrac{ \left[ 0; \ \boldsymbol{ \hat n } \right] }{ R } \left\{
\dfrac{ 4 \pi r^2 { \omega_s }^2 }{ 3 ( c \ - \boldsymbol{ \hat n } \cdot \textbf{v} )^2 }
\right\}
\ - \dfrac{ \left[ 0; \ \langle \boldsymbol{ \hat n } \times \boldsymbol{ \omega_s } \rangle \times \boldsymbol{ \omega_s } \right] }{ R } \left\{
\dfrac{ 4 \pi r^2 }{ 3 ( c \ - \boldsymbol{ \hat n } \cdot \textbf{v} )^2 } 
\right\}
\right.
\\
\\
\left.
\ + \dfrac{ \left[ 0; \ \textbf{v} \times \boldsymbol{ \omega_s } \right] }{ R^2 } \left\{
\dfrac{ 4 \pi r^2 }{ 3 ( c \ - \boldsymbol{ \hat n } \cdot \textbf{v} )^2 }
\right\}
\right]
\end{matrix}
\end{equation} \medskip

All terms like $ \boldsymbol{ \hat n } \cdot ( \boldsymbol{ \hat n } \times \boldsymbol{ \omega_s } ) $ vanish. Further, second term cancels-out with fourth and fifth, as $ ( \boldsymbol{ \hat n } \times \boldsymbol{ \omega_s } )^2 \ = { \omega_s }^2 \ - ( \boldsymbol{ \hat n } \cdot \boldsymbol{ \omega_s } )^2 $. The eighth term cancels-out with ninth. Lastly, the seventeenth and eighteenth terms cancel-out with nineteenth as $ ( \boldsymbol{ \hat n } \times \boldsymbol{ \omega_s } ) \times \boldsymbol{ \omega_s } \ = - { \omega_s }^2 \boldsymbol{ \hat n } \ + ( \boldsymbol{ \hat n } \cdot \boldsymbol{ \omega_s } ) \boldsymbol{ \omega_s } $. \bigskip

Thus, writing and rearranging only the surviving term out of equation (\ref{29}), we have :

\begin{equation} \label{30}
\begin{matrix}
A^\mu ( R^\mu ) \ = \lim_{ r_q \to \ 0 } \dfrac{ \rho }{ 4 \pi \epsilon_0 c^2 } \int_{ r= 0 }^{ r_q } 4 \pi \left[
{ \left[ c; \ \textbf{v} \right] } \left\{
\dfrac{ r^2 }{ R ( c \ - \boldsymbol{ \hat n } \cdot \textbf{v} ) }
\right\}
\right.
\\
\\
\left.
\ - \dfrac{ \left[ 0; \ \boldsymbol{ \hat n } \times \boldsymbol{ \omega_s } \right] }{ 3 } \left\{ 
\dfrac{ r^4 }{ R ( c \ - \boldsymbol{ \hat n } \cdot \textbf{v} )^2 }
\ + \dfrac{ r^4 ( \boldsymbol{ \hat n } \cdot \textbf{a}  ) }{ R ( c \ - \boldsymbol{ \hat n } \cdot \textbf{v} )^3 } 
\ + \dfrac{ r^4 ( c^2 \ - v^2 ) }{ R^2 ( c \ - \boldsymbol{ \hat n } \cdot \textbf{v} )^3 } 
\right\}
\right.
\\
\\
\left. 
\ + \dfrac{ \left[ 0; \ \textbf{v} \times \boldsymbol{ \omega_s } \right] }{ 3 } \left\{
\dfrac{ 4 \pi r^2 }{ R^2 ( c \ - \boldsymbol{ \hat n } \cdot \textbf{v}  )^2 }
\right\}
\right] dr
\end{matrix}
\end{equation} \medskip

Now, performing $ r $ integrals and re-arranging the resultant terms, we get:

\begin{equation} \label{31}
\begin{matrix}
A^\mu (R^\mu) \ = 
\lim_{ r_q \to \ 0 } \dfrac{ \rho }{ 4 \pi \epsilon_0 c^2 } \left( \dfrac{ 4 }{ 3 } \pi { r_q }^3 \right) \left[
\dfrac{ 1 }{ R } \left\{
\dfrac{ \left[ c; \ \textbf{v}  \right] }{ ( c \ - \boldsymbol{ \hat n } \cdot \textbf{v} ) }
\right. \right.
\\
\\
\left. \left.
\ - \dfrac{ r_q^2 \left[ 0; \ \boldsymbol{ \hat n } \times \boldsymbol{ \dot \omega_s } \right] }{ 5 ( c \ - \boldsymbol{ \hat n } \cdot \textbf{v} )^2 } 
\ - \dfrac{ r_q^2 ( \boldsymbol{ \hat n } \cdot \textbf{a} ) { \left[ 0; \ \boldsymbol{ \hat n } \times \boldsymbol{ \omega_s } \right] } }{ 5 ( c \ - \boldsymbol{ \hat n } \cdot \textbf{v} )^3 }
\right\}
\right.
\\
\\
\left.
\ + \dfrac{ 1 }{ R^2 } \left\{
\dfrac{ r_q^2 \left[ 0; \ \textbf{v}  \times \boldsymbol{ \omega_s } \right] }{ 5 ( c \ - \boldsymbol{ \hat n } \cdot \textbf{v} )^2 }
\ - \dfrac{ r_q^2 ( c^2 - v^2 ) \left[ 0; \ \boldsymbol{ \hat n } \times \boldsymbol{ \omega_s } \right] }{ 5 ( c \ - \boldsymbol{ \hat n } \cdot \textbf{v} )^3 }
\right\}
\right]
\end{matrix} 
\end{equation} \medskip

In the point-particle limit $ r_q \rightarrow 0 $, the total charge of the sphere $ 4/ 3 \pi { r_q }^3 \ \rho $ becomes $ q $, the specific charge of the point-particle. \bigskip

Further, in rest of the terms, quantity $ r_q^2 \boldsymbol{ \omega_s } $ can be represented as the particle's spin-angular-momentum $ \textbf{s} $. \bigskip

Spin-Angular-Momentum of a solid sphere is given by:

\begin{equation} \label{32}
\textbf{s} \ = \dfrac{ 2 }{ 5 } m { r_q }^2 \boldsymbol{ \omega_s }
\end{equation} \medskip

In point-particle limit $ r_q \rightarrow 0 $, $ \textbf{s} $ is known as 'Classical-spin' of the point-particle. \bigskip

Thus, using equation (\ref{31}) in point-particle limit, and writing equation (\ref{31}) for $ \Phi ( \textbf{R} , t ) $ and $ \textbf{A} ( \textbf{R} , t ) $  individually, we have:

\begin{equation} \label{33}
\Phi ( \textbf{R} , t ) \ = \dfrac{ q }{ 4 \pi \epsilon_0 R } \dfrac{ c }( c - \boldsymbol{ \hat n } \cdot \textbf{v} )
\end{equation} \medskip

and:

\begin{equation} \label{34}
\begin{matrix}
\textbf{A} ( \textbf{R} , t ) \ = \dfrac{ q }{ 4 \pi \epsilon_0 c } \left[
\dfrac{ 1 }{ R } \left\{
\dfrac{ \textbf{v} }{ ( c \ - \boldsymbol{ \hat n } \cdot \textbf{v} ) } 
\ - \dfrac{ ( \boldsymbol{ \hat n } \times { \boldsymbol{ \dot s } } ) }{ 2 m ( c \ - \boldsymbol{ \hat n } \cdot \textbf{v} )^2 }
\right. \right.
\\
\\
\left. \left.
\ - \dfrac{ ( \boldsymbol{ \hat n } \cdot \textbf{a} ) ( \boldsymbol{ \hat n } \times \textbf{s} ) }{ 2 m ( c \ - \boldsymbol{ \hat n } \cdot \textbf{v} )^3 }
\right\}
\ + \dfrac{ 1 }{ R^2 } \left\{
\dfrac{ ( \textbf{v} \times \textbf{s} ) }{ 2 m ( c \ - \boldsymbol{ \hat n } \cdot \textbf{v} )^2 }
\ - \dfrac{ ( c^2 - v^2 ) ( \boldsymbol{ \hat n } \times \textbf{s} ) }{ 2 ( c \ - \boldsymbol{ \hat n } \cdot \textbf{v} )^3 }
\right\}
\right]
\end{matrix}
\end{equation} \medskip

The scalar LW potential remains the same as in equation (\ref{2}).
However, the vector LW potential arrives with several additional spin-dependent terms. \medskip

\section{Interpretation:} \label{interp}

A comparison of equations (\ref{33}) and (\ref{34}) with (\ref{2}) reveals that the `original' LW potentials are, the zeroth order terms caused only by the motion of a  `non-spinning' point-charge. \bigskip

Further, a firsthand inspection of other terms of equation (\ref{34}) appears to yield the  following interpretation: \bigskip

The second term in equation (\ref{34}) contains time rate-of-change of spin. Thus, it is a transient term, referring to some mechanism by which the point-charge negotiates its own spin (orientation) with the system outside during transients, e.g. shifting orbits. \bigskip

The third term in equation (\ref{34}) indicates some sort of interaction between particle's acceleration and it's spin. \bigskip
The fourth term in equation (\ref{34}) resembles the Coriolis acceleration term. However, in this case, the particle's velocity interacts with its own spin rather than external revolving frame. We may call it “Self Coriolis Acceleration” term. \bigskip

The fifth, and the last, term, simply, is the 'Magnetic-Dipole' term that results from the multi-pole  expansion of $ \textbf{A} ( \textbf{R} , t ) $.  \cite{grif} \bigskip

However, the form of equation (\ref{34}) and the numerous terms do not seem to be in meaningful pattern and need further investigation to properly comprehend. \bigskip

Thus, to investigate further, we first formulate the equations (\ref{33}) and (\ref{34}) as sums of original LW potentials of equation (\ref{2}) and acquired spin-dependent additional (correction) terms, as:

\begin{equation} \label{35}
\left. 
\begin{matrix}
\Phi ( \textbf{R} , t ) & \ = \Phi_0 ( \textbf{R} , t ) \ + \delta \Phi ( \textbf{R} , t, \textbf{s} ) \\
{} & {} \\
\textbf{A} ( \textbf{R} , t ) & \ = \textbf{A}_0 ( \textbf{R} , t ) \ + \delta \textbf{A} ( \textbf{R} , t, \textbf{s} )
\end{matrix}
\\ \right\}
\end{equation} \medskip

While, $ \Phi_0 ( \textbf{R} , t ) $ and $ \textbf{A}_0 ( \textbf{R} , t ) $ being the original LW potentials given by equations (\ref{2})  ; i.e. due to a 'non-spinning' point-charge. Here '0' subscript indicates $ s \ = 0 $. \bigskip

Then, the spin-dependent correction terms are:

\begin{equation} \label{36}
\left.
\begin{matrix}
\delta \Phi ( \textbf{R} , t, \textbf{s} ) & \ = 0 \\
{}& {} \\
\delta \textbf{A} ( \textbf{R} , t, \textbf{s} ) & \ =
\dfrac{ q }{ 4 \pi \epsilon_0 c } \left[
\dfrac{ 1 }{ R } \left\{
\ - \dfrac{ ( \boldsymbol{ \hat n } \times \boldsymbol{ \dot s } ) }{ 2 m ( c \ - \boldsymbol{ \hat n } \cdot \textbf{v} )^2 }
\ - \dfrac{ ( \boldsymbol{ \hat n } \cdot \textbf{a} ) ( \boldsymbol{ \hat n } \times \textbf{s} ) }{ 2 m ( c \ - \boldsymbol{ \hat n } \cdot \textbf{v} )^3 }
\right\}
\right. \\
{}& \left.
\ + \dfrac{ 1 }{ R^2 } \left\{ 
\dfrac{ ( \textbf{v} \times \textbf{s} ) }{ 2 m ( c \ - \boldsymbol{ \hat n } \cdot \textbf{v} )^2 }
\ - \dfrac{ ( c^2 - v^2 ) ( \boldsymbol{ \hat n } \times \textbf{s} ) }{ 2 ( c \ - \boldsymbol{ \hat n } \cdot \textbf{v} )^3 }
\right\}
\right]
\end{matrix}
\\  \right\}
\end{equation} \medskip

The expression for $ \delta \textbf{A} ( \textbf{R} , t, \textbf{s} ) $ from equation (\ref{36}) may be re-written as:

\begin{equation} \label{37}
\begin{matrix}
\delta \textbf{A} ( \textbf{R} , t, \textbf{s} ) \ = & 
\dfrac{ q }{ 8 \pi \epsilon_0 m c } \left[
\ - \dfrac{ ( \textbf{R} \times \boldsymbol{ \dot s } ) }{ ( R c \ - \textbf{R}  \cdot \textbf{v} )^2 }
\ - \dfrac{ ( \textbf{R} \cdot \textbf{a} ) ( \textbf{R}  \times \textbf{s} ) }{ ( R c \ - \textbf{R} \cdot \textbf{v} )^3 }
\right. \\ 
{} & \left.
\ + \dfrac{ ( \textbf{v} \times \textbf{s} ) }{ ( R c \ - \textbf{R} \cdot \textbf{v} )^2 }
\ - \dfrac{ ( c^2 - v^2 ) ( \textbf{R} \times \textbf{s} ) }{ ( R c \ - \textbf{R} \cdot \textbf{v} )^3 }
\right]
\end{matrix}
\end{equation} \medskip

The various terms may be re-arranged as:

\begin{equation} \label{38}
\begin{matrix}
\delta \textbf{A} ( \textbf{r} , t , \textbf{s} ) & \ =
\dfrac{ q }{ 8 \pi \epsilon_0 m c } \left[\
\left( \dfrac{ - \textbf{R} }{ R c \ - \textbf{R} \cdot \textbf{v} } \right) \times 
\left( \dfrac{ \boldsymbol{ \dot s } }{ R c \ - \textbf{R} \cdot \textbf{v} } \right)
\right. \\
{} & {} \\
{}& \left.
\ +   \left\{ \textbf{v} 
\ + ( c^2 \ - v^2 \ + \textbf{R} \cdot \textbf{a} ) \left( \dfrac{ - \textbf{R} }{ R c \ - \textbf{R} \cdot \textbf{v} } \right) 
\right\}
\times \left\{ 
\dfrac{ \textbf{s} }{ ( R c \ - \textbf{R} \cdot \textbf{v} )^2 } 
\right\}
\right]
\end{matrix}
\end{equation} \medskip

Using expression for gradient of retarded-time $ \boldsymbol{ \nabla t_{ret} } $ \cite{grif}:

\begin{equation} \label{39}
\boldsymbol{ \nabla } t_{ret} \ = \dfrac{ - \textbf{R} }{ R c \ - \textbf{R} \cdot \textbf{v} }
\end{equation} \medskip

And \cite{grif}:

\begin{equation} \label{40}
\boldsymbol{ \nabla } ( R c \ - \textbf{R} \cdot \textbf{v} ) \ = 
( v^2 \ - c^2 \ - \textbf{R} \cdot \textbf{a} ) \boldsymbol{ \nabla } t_{ret} \ - \textbf{v} 
\end{equation} \medskip

Equation (\ref{38}) may be written as:

\begin{equation} \label{41}
\delta \textbf{A} ( \textbf{R} , t, \textbf{s} ) \ = \dfrac{ q }{ 8 \pi \epsilon_0 m c } \left[
\dfrac{ \boldsymbol{ \nabla } t_{ret} \times \boldsymbol{ \dot s } }{ ( R c \ - \textbf{R} \cdot \textbf{v} ) }
\ - \dfrac{ \boldsymbol{ \nabla } ( R c \ - \textbf{R} \cdot \textbf{v} ) }{ ( R c \ - \textbf{R} \cdot \textbf{v} )^2 } \times \textbf{s} 
\right]
\end{equation} \medskip

Now \cite{grif}:

\begin{equation} \label{42}
\boldsymbol{ \nabla } \times \textbf{s} \ = - \left( \dfrac{ d \textbf{s} }{ d t_{ret} } \right) \times \boldsymbol{ \nabla } t_{ret} \ = \boldsymbol{ \nabla } t_{ret} \times { \boldsymbol{ \dot s} }
\end{equation} \medskip

And \cite{grif}:

\begin{equation} \label{43}
\boldsymbol{ \nabla } \left( \dfrac{ 1 }{ R c \ - \textbf{R} \cdot \textbf{v} } \right) \ = 
\dfrac{ - \boldsymbol{ \nabla } ( R c \ - \textbf{R} \cdot \textbf{v} ) }{ ( R c \ - \textbf{R} \cdot \textbf{v} )^2 }
\end{equation} \medskip

Making use of equations (\ref{42}) and (\ref{43}), equation (\ref{41}) becomes:

\begin{equation} \label{44}
\delta \textbf{A} ( \textbf{R} , t, \textbf{s} ) \ = \dfrac{ q }{ 8 \pi \epsilon_0 m c } \left[
\dfrac{ \boldsymbol{ \nabla } \times \textbf{s} }{ ( R c \ - \textbf{R} \cdot \textbf{v} ) }
\ + \boldsymbol{ \nabla } \left( \dfrac{ 1 }{ R c \ - \textbf{R} \cdot \textbf{v} } \right) \times \textbf{s}
\right]
\end{equation} \medskip

Which finally simplifies to:

\begin{equation} \label{45}
\delta \textbf{A} ( \textbf{R} , t, \textbf{s} ) \ = \dfrac{ q }{ 8 \pi \epsilon_0 m c } \boldsymbol{ \nabla } \times \left(
\dfrac{ \textbf{s} }{ R c \ - \textbf{R} \cdot \textbf{v} }
\right)
\end{equation} \medskip

Thus, summing up equation (\ref{44}) with (\ref{35}), and along with (\ref{36}) and (\ref{2}), we have the final expression for LW potentials:

\begin{equation} \label{46}
\left.
\begin{matrix}
\Phi ( \textbf{R} , t ) & \ = &
\dfrac{ q c }{ 4 \pi \epsilon_0 } \left[
\dfrac{ 1 }{ ( R c \ - \textbf{R} \cdot \textbf{v} ) }
\right] \\
{} & {} & {} \\
\textbf{A} ( \textbf{R} , t ) & \ = &
\dfrac{ q }{ 4 \pi \epsilon_0 c } \left[
\dfrac{ \textbf{v} }{ ( R c \ - \textbf{R} \cdot \textbf{v} ) }
\ + \dfrac{ 1 }{ 2 m } \boldsymbol{ \nabla } \times  \left( \dfrac{ \textbf{s} }{ R c \ - \textbf{R} \cdot \textbf{v} }
\right)
\right]
\end{matrix}
\ \  \right\}
\end{equation} \medskip

As the quantity $  R c \ - \textbf{R} \cdot \textbf{v} \ = - R_\mu u^\mu $ in denominator represents apparent geometry due to 'retardation', the vector $ { \textbf{s} } / ( R c \ - \textbf{R} \cdot \textbf{v} ) $ may be referred to as “apparent spin”: the classical-spin that is 'perceived' by a distant observer. \bigskip

Therefore, the correction terms obtained in $ \textbf{A} ( \textbf{R} , t ) $ are collectively proportional to the 'rotation' (curl) of the apparent spin. \bigskip

However, in view of equations (\ref{46}) there is an another important observation, that  the previously known coupling between LW scalar and vector potentials would no longer hold; i.e.:

\begin{equation} \label{47}
\textbf{A} ( \textbf{R} , t ) \ \neq \ \dfrac{ \textbf{v} }{ c^2 } \Phi ( \textbf{R} , t )
\end{equation} \medskip

Rather, another intuitive relation between LW scalar and vector potentials appears:

\begin{equation} \label{48}
\boldsymbol{ \hat{n} } \cdot \textbf{A} ( \textbf{R} , t ) \ = \ \dfrac{ ( \boldsymbol{ \hat{n} } \cdot \textbf{v} ) }{ c^2 } \Phi ( \textbf{R} , t )
\end{equation} \medskip

Or, in other words, only the ’parallel’ components of $ \textbf{A} ( \textbf{R} , t ) $ and $ \textbf{v} $, in the direction of faraway observer satisfy:

\begin{equation} \label{49}
A_{ {} \parallel {} } ( \textbf{R} , t ) \ = \ \dfrac{ v_{ {} \parallel {} } }{ c^2 } \ \Phi ( \textbf{R} , t )
\end{equation} \medskip

In upcoming section we shall check the validity of equations (\ref{46}). \medskip

\section{Validation:} \label{valid}

Li\'enard-Wiechert potentials are Retarded-Potentials, which being in Lorenz Gauge, satisfy Lorenz Gauge condition \cite{grif} \cite{jak}:

\begin{equation} \label{50}
\boldsymbol{ \nabla } \cdot  \textbf{A} ( \textbf{R} , t ) \ = - \dfrac{ 1 }{ c^2 } \dfrac{ \partial \Phi ( \textbf{R} , t ) }{ \partial t }
\end{equation} \medskip

In addition, they also satisfy the homogeneous wave-equations in vacuum \cite{grif} \cite{jak}:

\begin{equation} \label{51}
\left. 
\begin{matrix}
{ \square }^2 \Phi ( \textbf{R} , t ) \ = & { \nabla }^2 \Phi ( \textbf{R} , t ) \ - \dfrac{ 1 }{ c^2 } \dfrac{ \partial^2 \Phi ( \textbf{R} , t ) }{ \partial t^2 } \ = & 0 \\
{} & {} & {} \\
{ \square }^2 \textbf{A} ( \textbf{R} , t ) \ = &  { \nabla }^2 \textbf{A} ( \textbf{R} , t ) \ - \dfrac{ 1 }{ c^2 } \dfrac{ \partial^2 \textbf{A} ( \textbf{R} , t ) }{ \partial t^2 } \ = & 0
\end{matrix}
\ \  \right\}
\end{equation} \medskip

If the derived LW potentials of equations (\ref{46}) are valid, they must also meet the criteria of equations (\ref{50}) and (\ref{51}). \bigskip

The part of equations (\ref{35}) , $ \Phi_0 ( \textbf{R} , t ) $ and $ \textbf{A} _0 ( \textbf{r} , t ) $ are simply the original LW potentials, thus they already satisfy (\ref{50}) and (\ref{51}). \bigskip

Further, as there is no correction term in $ \Phi ( \textbf{R} , t ) $ (i.e. $ \delta \Phi ( \textbf{R} , t , \textbf{s} ) \ = 0 $ ) according to equation (46), the only test remains is for $ \delta \textbf{A} ( \textbf{R} , t , \textbf{s} ) $, which must satisfy equations (\ref{50}) and (\ref{51}). \bigskip

As from equation (\ref{36}), $ \delta \textbf{A} ( \textbf{R} , t , \textbf{s} ) $ being the curl of a vector; its divergence vanishes. And, since, $ \delta \Phi ( \textbf{R} , t , \textbf{s} ) \ = 0 $, it immediately satisfies the Lorenz gauge condition:

\begin{equation} \label{52}
\boldsymbol{ \nabla } \cdot  \delta \textbf{A} ( \textbf{R} , t ) \ = 0 \ = - \dfrac{ 1 }{ c^2 } \dfrac{ \partial \delta \Phi ( \textbf{R} , t ) }{ \partial t }
\end{equation} \medskip

Now, writing the L.H.S. Of the vacuum homogeneous wave-equation for $ \delta \textbf{A} ( \textbf{R} , t , \textbf{s} ) $ and  using equation (\ref{45}), we have:

\begin{equation} \label{53}
{ \nabla }^2 \delta \textbf{A} \ - \dfrac{ 1 }{ c^2 } \dfrac{ \partial ^2 \delta \textbf{A} }{ \partial t^2 } \ = 
\dfrac{ q }{ 8 \pi \epsilon_0 m c }
\left \langle { \nabla }^2 \ - \dfrac{ 1 }{ c^2 } \dfrac{ \partial ^2 }{ \partial t^2 } \right \rangle 
\left[ 
\boldsymbol{ \nabla } \times \left( \dfrac{ \textbf{s} }{ R c \ - \textbf{R} \cdot \textbf{v} } \right)
\right]
\end{equation} \medskip

Curl and Laplacian operators are orthogonal and thus are interchangeable. Also the curl operator, being a space-derivative, is independent of time; so the curl operator and time-derivative are also interchangeable. Therefore, equation (\ref{53}) may be written as:

\begin{equation} \label{54}
\nabla^2 \delta \textbf{A} \ - \dfrac{ 1 }{ c^2 } \dfrac{ \partial ^2 \delta \textbf{A} }{ \partial t^2 } \ =
\dfrac{ q }{ 8 \pi \epsilon_0 m c }
\boldsymbol{ \nabla } \times \left[ 
\left \langle \nabla^2 \ - \dfrac{ 1 }{ c^2 } \dfrac{ \partial ^2}{ \partial t^2 } \right \rangle
\left( \dfrac{ \textbf{s} }{ R c \ - \textbf{R} \cdot \textbf{v} } \right) 
\right]
\end{equation} \medskip

Equation (\ref{54}) says that if the vector $ { \textbf{s} }/ ( R c \ - \textbf{R} \cdot \textbf{v} ) $ satisfies the vacuum homogeneous wave-equation, $ \delta \textbf{A} ( \textbf{r} , t , \textbf{s} ) $ would also  satisfy it. \bigskip

Thus, for $ \delta \textbf{A} ( \textbf{r} , t , \textbf{s} ) $ to be valid:

\begin{equation} \label{55}
\nabla^2 \left( \dfrac{ \textbf{s} }{ R c \ - \textbf{R} \cdot \textbf{v} } \right) \ - 
\dfrac{ 1 }{ c^2 } \dfrac{ \partial ^2}{ \partial t^2 } \left( \dfrac{ \textbf{s} }{ R c \ - \textbf{R}  \cdot \textbf{v} } \right) \ \equiv 0
\end{equation} \medskip

Solving the L.H.S. of equation (\ref{55}):

\begin{equation} \label{56}
\begin{matrix}
\left \langle \nabla^2 \ - \dfrac{ 1 }{ c^2 } \dfrac{ \partial ^2 }{ \partial t^2 } \right \rangle \left( \dfrac{ \textbf{s} }{ R c \ - \textbf{R} \cdot \textbf{v} } \right) \ =
\dfrac{ 1 }{ ( R c \ - \textbf{R} \cdot \textbf{v} ) } \left( \nabla^2 \textbf{s} \ - \dfrac{ 1 }{ c^2 } \dfrac{ \partial ^2 \textbf{s} }{ \partial t^2 } \right)
\\
\\ 
\ +  \textbf{s} \left \langle \nabla^2 \ - \dfrac{ 1 }{ c^2 } \dfrac{ \partial ^2 }{ \partial t^2 } \right \rangle \left( \dfrac{ 1 }{ R c \ - \textbf{R}  \cdot \textbf{v} } \right)
\\
\\ 
\ + 2 \left( \dfrac{ \partial \textbf{s} }{ \partial t_{ret} } \right) \left\{
( \boldsymbol{ \nabla } t_{ret} ) \cdot \boldsymbol{ \nabla } \left( \dfrac{ 1 }{ R c \ - \textbf{R} \cdot \textbf{v} } \right)
\ - \dfrac{ 1 }{ c^2 } \left( \dfrac{ \partial t_{ret} }{ \partial t } \right) \dfrac{ \partial }{ \partial t } \left( \dfrac{ 1 }{ R c \ - \textbf{R} \cdot \textbf{v} } \right)
\right\}
\end{matrix}
\end{equation} \medskip

The middle term vanishes, because the function $ 1 / ( R c \ - \textbf{R} \cdot \textbf{v} ) $ defines the geometry of original Scalar LW potential of equation (\ref{2}); thus satisfies homogeneous vacuum wave equation:

\begin{equation} \label{57}
\begin{matrix}
\left[ \nabla^2 \ - \dfrac{ 1 }{ c^2 } \dfrac{ \partial ^2 }{ \partial t^2 } \right] \Phi( \textbf{R} , t) & \ = &
\left[ \nabla^2 \ - \dfrac{ 1 }{ c^2 } \dfrac{ \partial ^2 }{ \partial t^2 } \right] \left[
\dfrac{ q }{ 4 \pi \epsilon_0 } \dfrac{ c }{ ( Rc - \textbf{R}  \cdot \textbf{v} ) }
\right] \\
{} & {} & {} \\
{} & \ = &
\dfrac{ q c }{ 4 \pi \epsilon_0 } \left[ \nabla^2 \ - \dfrac{ 1 }{ c^2 } \dfrac{ \partial ^2 }{ \partial t^2 } \right] \left( \dfrac{ 1 }{ Rc - \textbf{R} \cdot \textbf{v} } \right) \ = 0
\end{matrix}
\end{equation} \medskip

Further, it may be shown via evaluation of various space and time derivatives that:

\begin{equation} \label{58}
\nabla^2 \textbf{s} \ - \dfrac{ 1 }{ c^2 } \dfrac{ \partial ^2 \textbf{s} }{ \partial t^2 } \ = \dfrac{ - 2 }{ ( R c - \textbf{R}  \cdot \textbf{v} ) } \dfrac{ \partial \textbf{s} }{ \partial t_{ret} }
\end{equation} \medskip

And:

\begin{equation} \label{59}
( \boldsymbol{ \nabla } t_{ret} ) \cdot \boldsymbol{ \nabla } \left( \dfrac{ 1 }{ R c \ - \textbf{R} \cdot \textbf{v} } \right)
\ - \dfrac{ 1 }{ c^2 } \left( \dfrac{ \partial t_{ret} }{ \partial t } \right) \dfrac{ \partial }{ \partial t } \left( \dfrac{ 1 }{ R c \ - \textbf{R} \cdot \textbf{v} } \right) 
\ = \dfrac{ 1 }{ ( R c \ - \textbf{R} \cdot \textbf{v} )^2 }
\end{equation} \medskip

Thus, with use of equations (\ref{58}), (\ref{59}) and (\ref{57}) in (\ref{56}), we get:

\begin{equation} \label{60}
\left \langle \nabla^2 \ - \dfrac{ 1 }{ c^2 } \dfrac{ \partial ^2 }{ \partial t^2 } \right \rangle \left( \dfrac{ \textbf{s} }{ R c \ - \textbf{R} \cdot \textbf{v} } \right) \ = 0
\end{equation} \medskip

Therefore, from equation (\ref{60}) and (\ref{54}), it is proved that $ \delta \textbf{A} ( \textbf{R} , t , \textbf{s} ) $ satisfies vacuum homogeneous wave-equation. \bigskip

Thus, the derived expressions (\ref{34}) and (\ref{46}) for LW potentials due to a spinning and non-relativistically moving point-charge, are valid in the context of classical electrodynamics.  \medskip

\section{Conclusion:} \label{conclu}

In this paper LW potentials due to a spinning point-charge in arbitrary motion are derived, beginning with a spinning and non-relativistically moving charged-sphere and subsequently reducing its dimension to the `point-charge' limit. \bigskip

Though, expression for LW scalar-potential stays the same, several, spin-dependent correction terms are discovered in the expression for the  LW vector-potential (\ref{34}) that coalesce into a single correction term in equation (\ref{46}), which is proportional to the `rotation' of apparent classical-spin. \bigskip

However, equation (\ref{47}) also reveals that the previously known $ \textbf{v}  / {c^2} $ coupling between the scalar and vector LW potentials is not valid generally; but only for the components that lie in the observer's direction. \bigskip

The importance of the acquired correction term(s) needs to be further studied by deriving electromagnetic-fields and radiated power from them. Though, the same is not in this paper's scope. \bigskip

We believe that these `spin' dependent terms, in LW vector-potential may alter (or extend) the physics of EM fields and radiation (from a point-charge); nevertheless, further investigation is required. \bigskip

One observation is quite intriguing: in Quantum-Mechanics spin of elementary charged-particles are seen as $ h / 2 $ (or odd multiple of $ h / 2 $); while $ h $ being the Plank's constant, which bears the dimensions of angular-momentum. Whereas, in our expressions (\ref{34}) and (\ref{46}), additional terms in LW potentials are proportional to $ \textbf{s} / 2 $. Thus we believe the work of this paper will  be crucial in establishing some link between the Classical and Quantum Physics. \bigskip

Lastly, in this paper, the LW potentials of equations (\ref{46}) are the outcomes of series approximation in the non-relativistic case; where only the terms up to $ r^2 $ are examined. However, to acquire a better understanding of the classical-spin dependence of potentials (and fields), complete analytical solution needs to be evaluated. \bigskip

\textbf{ Data availability: } No data has been analyzed or generated, because, our work is a purely theoritical and mathematical approach. \medskip

\printbibliography

\end{document}